\documentclass[aps,pra,reprint,twocolumn,floatfix,superscriptaddress,longbibliography]{revtex4-1}
\usepackage[T1]{fontenc} 
\usepackage{mathtools}
\usepackage{standalone}
\usepackage{amsthm}
\usepackage{lipsum}
\usepackage{amsfonts, amsmath, amssymb}
\usepackage{graphicx}
\usepackage{dcolumn}
\usepackage{bm}
\usepackage{dsfont}
\usepackage[normalem]{ulem}
\usepackage{color}
\usepackage[utf8]{inputenc}
\usepackage[colorlinks,citecolor=blue,linkcolor=blue,urlcolor=blue]{hyperref}
\usepackage{tikz}
\usepackage{braket}
\usepackage{physics}
\usepackage{color}
\usepackage{soul}
\usepackage{mathtools}
\usepackage{float}
\usepackage{esvect}
\usepackage{subfigure}
\usepackage{stmaryrd} 
\usepackage{enumerate}

\newcommand{\dcal}{\mathcal{D}}
\newcommand{\bea}{\begin{equation}\begin{aligned}}
\newcommand{\eea}{\end{aligned}\end{equation}}
\newcommand{\be}{\begin{equation}}
\newcommand{\ee}{\end{equation}}

\newcommand{\ntrain}{N_{\mathrm{train}}}
\newcommand{\ntest}{{N_{\mathrm{test}}}}
\newcommand{\npulse}{{N_{\mathrm{pulse}}}}

\newcommand{\meanq}[1]{\langle #1 \rangle}
\newcommand{\expect}[1]{\mathbb{E}\left[ #1\right]}
\newcommand{\expectt}[2]{\mathbb{E}_{#1}\left[ #2\right]}

\renewcommand{\(}{\left(}
\renewcommand{\)}{\right)}
\renewcommand{\[}{\left[}
\renewcommand{\]}{\right]}

\def\rmi{{\rm {i}}}

\def\tr{{\rm{Tr}}}
\newcommand{\sigmahat}{\hat{\sigma}}

\newcommand{\Hhat}{\hat{H}}
\newcommand{\rhohat}{\hat{\rho}}

\newcommand{\id}{\mathds{1}}

\newcommand{\hsx}{\hat{\sigma}_x}
\newcommand{\hsy}{\hat{\sigma}_y}
\newcommand{\hsz}{\hat{\sigma}_z}

\newtheorem{theorem}{Theorem}[section]

\theoremstyle{definition}

\theoremstyle{remark}

  {\left\lbrace\begin{array}{@{}l@{}}}%
  {\end{array}\right.}

\makeatletter
\newcommand{\phantomlabel}[2]{
    \protected@write\@auxout{}{
        \string\newlabel{#2}{
            {\@currentlabel#1}{\thepage}
            {\@currentlabel#1}{#2}{}
        }
    }
    \hypertarget{#2}{}
}
\makeatother

\begin{document}

\title{Noisy Quantum Kernel Machines}

\author{Valentin Heyraud}
\author{Zejian Li}
\author{Zakari Denis}
\author{Alexandre Le Boit\'{e}}
\author{Cristiano Ciuti}
\affiliation{Universit\'{e} Paris Cit\'{e}, CNRS, Laboratoire Mat\'{e}riaux et Ph\'{e}nom\`{e}nes Quantiques (MPQ), F-75013 Paris, France}

\begin{abstract}

In the noisy intermediate-scale quantum era, an important goal is the conception of implementable algorithms that exploit the rich dynamics of quantum systems and the high dimensionality of the underlying Hilbert spaces to perform tasks while prescinding from noise-proof physical systems. An emerging class of quantum learning machines is that based on the paradigm of quantum kernels. Here, we study how dissipation and decoherence affect their performance. We address this issue by investigating the expressivity and the generalization capacity of these models within the framework of kernel theory. We introduce and study the effective kernel rank, a figure of merit that quantifies the number of independent features a noisy quantum kernel is able to extract from input data. Moreover, we derive an upper bound on the generalization error of the model that involves the average purity of the encoded states. Thereby we show that decoherence and dissipation can be seen as an implicit regularization for the quantum kernel machines. As an illustrative example, we report exact finite-size simulations of machines based on chains of driven-dissipative quantum spins to perform a classification task, where the input data are encoded into the driving fields and the quantum physical system is fixed. We determine how the performance of noisy kernel machines scales with the number of nodes (chain sites) as a function of decoherence and examine the effect of imperfect measurements. 
 
\end{abstract}

\date{\today}
\maketitle

\section{Introduction}

In recent years, machine learning has blossomed in a wide variety of fields and delivered a large number of applications driven by the achievements of the ever-developing field of artificial neural networks, particularly those presenting deep architectures~\cite{lecun2015, goodfellow2016}. Neural networks build predictions upon processing the input data of interest through a series of parametrized nonlinear transformations, whose (typically numerous) parameters are determined by training. This optimization procedure most often consists in minimizing a task-dependent loss function that quantifies the error of the parametrized model over a training dataset. This is most often implemented via software executed on standard computers.
As a result, the growing demand for computational resources and energy for training such deep architectures on ever-increasing amounts of data makes its long-term sustainability uncertain~\cite{strubell2019}. In this context, devolving computationally demanding tasks to machine-learning devices with suitable physical systems acting as hardware is emerging as a relevant alternative. However, while the neural-network sequential architecture is well suited for software implementations on standard computers, the great number of parameters to be tuned during training remains in practice an obstacle to physical implementations. A simpler alternative approach is provided by the category of ``shallow models'', such as reservoir-computing~\cite{tanaka2019} or extreme learning machines~\cite{huang2006}, which have led to physical proposals~\cite{opala2019, denis2022} and experimental realizations \cite{ballarini2020, pierangeli21}. In such machines, the input data are encoded in the dynamics of a physical system and the associated predictions are obtained by considering a linear combination of measured observables, weighted by a set of trainable parameters to be optimized by training. Importantly, this is done while keeping the parameters of the physical system fixed, hence requiring hardly any degree of control over the system. Kernel machines, whose trial functions can be represented in terms of positive semi-definite and symmetric kernel functions~\cite{hofmann2008}, belong to this category. More generally, kernel theory has proved to be a very useful tool to understand a wide range of machine-learning algorithms. Recently, a close connection between kernel machines and deep neural networks in the infinite width limit has been established~\cite{jacot2018}, further extending the relevance of these methods.

In parallel to the advent of quantum information, the last decade has also witnessed a growing interest in the emerging field of quantum machine learning~\cite{biamonte2017, dunjko2018}, a research domain that explores the potential advantages of quantum systems for machine-learning applications. Due to the success of deep neural-network algorithms, a large amount of work in this field has been devoted to finding quantum analogs to neural-network models~\cite{schuld2014}, and more generally to find brain-inspired algorithms to be implemented on quantum devices~\cite{markovic2020}. Parametrized quantum circuits used as trainable ansätze~\cite{benedetti2019} appeared as natural candidates for such a generalization. These models, often called quantum neural networks~\cite{farhi2018a}, are among the most studied quantum machine-learning models and significant progress has been achieved in the comprehension of their properties. Analogously to classical systems, quantum ``shallow" machines have also been put forward, such as those based on quantum reservoir-computing, extreme learning and quantum kernels \cite{park2020, schuld2019, lloyd2020, hubregtsen2021, schuld2021, kusumoto2021, liu2021, wu2021, kubler2021, shaydulin2021, wang2021, bartkiewicz2020}, where the physical system (the network) is fixed and the optimization concerns only a linear map acting on measured outcomes.

\begin{figure*}[t!]
    \centering
    \includegraphics[width=0.9\textwidth]{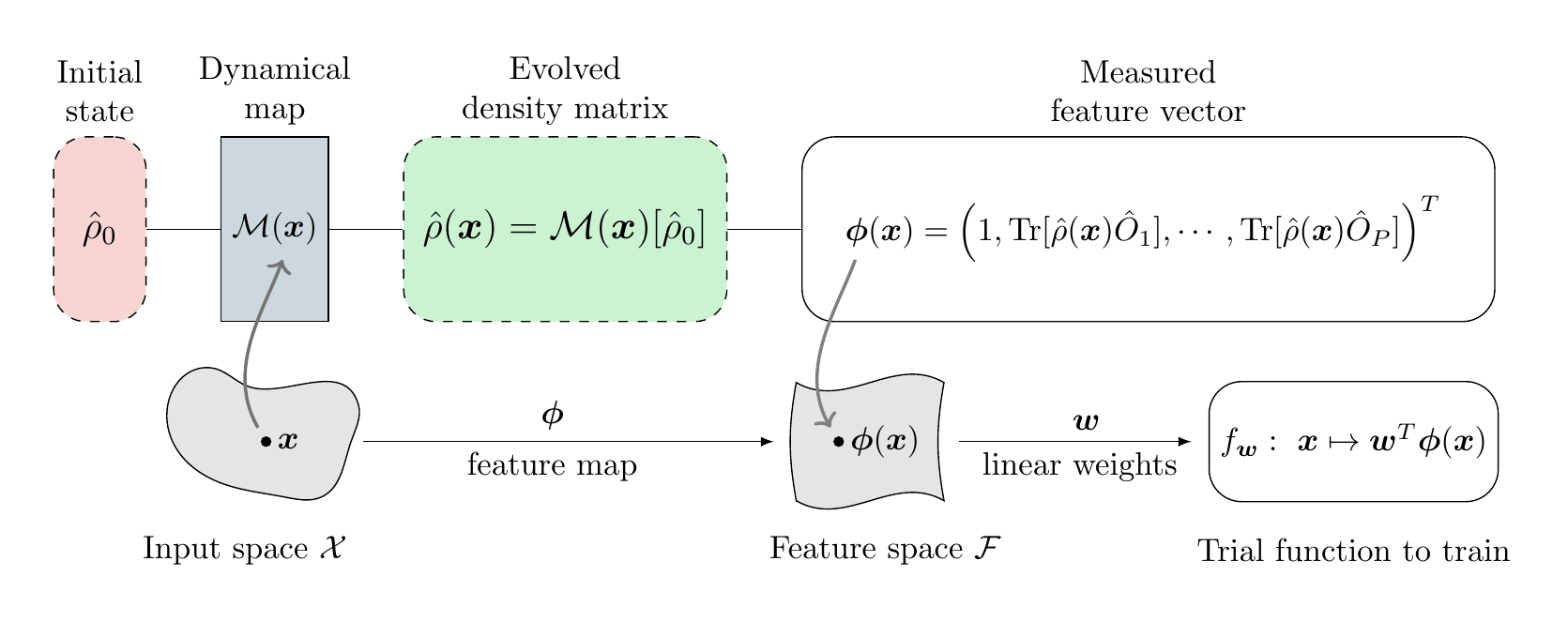}
    \caption{Scheme of a noisy quantum kernel machine. An element $\vb*{x}$ of the input space $\mathcal{X}$ is encoded into a density matrix $\rhohat(\vb*{x})$ obtained by evolving in time a fixed initial state described by the density matrix $\rhohat_0$ [see Fig. \ref{fig:enc_mnist} for a specific example of the encoding process described by the evolution map $\mathcal{M}(\vb*{x})$]. The measured features are represented by a vector of observables  $\vb*{\phi}(\vb*{x})$ (with an added $1$ corresponding to the unity operator as first element to create an offset term) that belongs to the feature space $\mathcal{F}$. The trial function is obtained by applying a linear transformation to the feature vector (depending nonlinearly on $\vb*{x}$) with a vector of weights $\vb*{w}$ that is optimized via the training procedure described in the main text.}
    \label{fig:dec_sch}
\end{figure*}

Most often, quantum machine-learning investigations have been focusing on isolated quantum systems with unitary dynamics. At present, however, we are in the so-called noisy intermediate-scale quantum era~\cite{preskill2018}: most quantum devices within practical reach are subject to a significant degree of dissipation and/or decoherence. An important problem is therefore to understand the impact of realistic noise on quantum machine-learning settings. The literature on the subject is yet in its very infancy. For time-dependent tasks, one study on quantum reservoir computing suggested that dissipation increases the processing capacity and the non-linearity of the embedding, at the price of a reduced memory capacity of the system~\cite{fujii2017}. An advantageous scaling of the performance of a quantum reservoir-computing scheme, as compared to its classical counterpart, was recently reported~\cite{xu2021}. However, a systematic study of the dissipation and decoherence on quantum machine-learning models is missing. In particular, to the best of our knowledge, no investigation has explored its role on the important class of quantum kernel machines. 

In this article, we investigate the use of open quantum systems as noisy quantum kernel machines. Within the formalism of kernel theory, we show how the expressive power and generalization capacity of the corresponding nonlinear feature maps are controlled by both the dissipation and decoherence affecting the system as well as the level of experimental uncertainty on the physical measurements. We introduce and study the effective kernel rank to quantify the effective number of independent features a noisy quantum kernel is able to extract from the input data. Moreover, we derive an upper bound on the generalization error of the model that involves the average purity of the encoded states. As an illustrative example, we simulate noisy quantum kernel machines implemented via driven-dissipative chains of spins.  We provide a comprehensive study of the performance of noisy quantum kernel machines, showing how they scale with the number of network nodes (chain sites) and the degree of dissipation and decoherence.

The paper is organized as follows. In section \ref{section:setup_and_model}, we describe the general scheme for encoding the input data into the quantum system dynamics and decoding the output through measurements.  In section \ref{section:intro_quantum_kernel_and_dissipation}, we analyze the noisy quantum kernel machine within the kernel-theory framework. In particular we study the link between the kernel spectrum and important properties of machine-learning models, such as the expressive power and the generalization capacity. We introduce and study the effective kernel rank. Within a statistical-learning approach, we provide an upper-bound on the generalization error for noisy quantum kernels. In section \ref{section:driven_dissipative_spins_model}, we describe a class of noisy quantum kernel machines based on driven-dissipative chains of spins. We report a comprehensive study of the dependence of the performance metrics on the system size and noise for this class of models in section \ref{section_numerical_results}. Finally, conclusions and perspectives are drawn in section \ref{section:conclusion}. The most technical details are reported in Appendices \ref{appendix:A_expressivity_and_generalization_dissipative_quantum_kernel}, \ref{appendix:B_intercept_and_kernel_centering} and \ref{appendix:C_time_multi_expressivity}.

\section{General scheme}

\label{section:setup_and_model}

The objective of supervised learning is to approximate a causal relation between elements $\vb*{x}$ of an input set $\mathcal{X}$ and some target quantities $y\in\mathcal{Y}$, based upon a set of known training examples $\mathcal{S} = \{(\vb*{x}_i, y_i) \,|\, i=1,\dots,\ntrain \}$. The input features are considered as independent realizations of a random variable following a probability distribution $p(x)$ on $\mathcal{X}$. In the following, we denote $\expectt{p}{f(\vb*{x})}$ the expectation value of a quantity $f(\vb*{x})$ over the distribution $p$ \footnote{In the case of multivariate functions the variables against which the expectation is taken is indicated in subscript.}. We also define the corresponding centered quantity as
\bea
\delta f(\vb*{x}) = f(\vb*{x}) - \expectt{p}{f}.
\eea
Upon assuming the inputs and target quantities are related according to an unknown ground-truth function $y_i = y(\vb*{x}_i)$, we aim to approximate it using a trial function $f$ parametrized by $\vb*{w}$, to be optimized using the training set $\mathcal{S}$. 
The specific form of $f$ depends on the considered model architecture. 
In this paper, we  describe noisy quantum kernel machines exploiting the dynamics of open quantum systems to generate such a trial function.
This scheme is summarized pictorially in Fig.~\ref{fig:dec_sch}.

\subsection{Encoding on the quantum system}
\label{section:system_and_encoding_method}

Let us consider a system initially prepared in a state $\hat{\rho}_0$. For each element of the input space, represented by a vector $\vb*{x} \in \mathcal{X}$, a procedure can be defined to encode it into the non-unitary dynamics of a generic open quantum system.
As will be shown in Sec.~\ref{section:driven_dissipative_spins_model}, this can be achieved, for instance, by encoding the input vector in a proper modulation of the driving fields acting on the system. 

We consider the dynamics of the open quantum system to be described by a Lindblad master equation~\cite{breuer2007} of the form:
\bea
 \dfrac{\partial\rhohat} {\partial t} = -\frac{\rmi}{\hbar}[\Hhat,\rhohat ]+ \sum_{j=1}^N \gamma_j \dcal(\hat{A}^j)[\rhohat ],
\label{eq:mastereq}
\eea
where $\gamma_j$ is the relaxation rate at site $i$ and the dissipator $\dcal(\hat{A})$ denotes the superoperator
\bea
    \dcal(\hat{A})\[\rhohat\] = \hat{A}^{\dagger}\rhohat\hat{A} - \frac{1}{2}\{ \hat{A}^{\dagger}\hat{A},\rhohat \}.
\eea
Note that the Lindblad operator $\hat{A}^j$ depends on the considered system-bath interaction. The master equation describes the evolution from an initial density matrix into a final density matrix:
\bea
\rhohat ( \vb*{x},t ) = {\mathcal M (\vb*{x},t})[\rhohat_{0}],
\eea
where the completely-positive trace-preserving map ${\mathcal M (\vb*{x},t})$ is the propagator of the Lindblad master equation capturing the non-unitary evolution of $\hat{\rho}_0$. It depends on $\vb*{x}$ via the encoding procedure: If the input is encoded in driving fields, as we will consider later, the Hamiltonian, and consequently the density matrix at any time, bears a dependence on the input. In principle, one could also encode the input into a modulation of the loss rates, although we will not treat this case here. In what follows, when considering a fixed final time $t_f$ for the time-evolution, we denote $\mathcal{M}(\vb*{x})=\mathcal{M}(\vb*{x},t_f)$ and $\rhohat(\vb*{x}) = \rhohat(\vb*{x},t_f)$ to simplify the notation.
\subsection{Decoding through measurements}
\label{section:decoding_methods}

At time $t_f$, after the encoding procedure, we extract the processed information by performing a set of measurements of the system. Given the density matrix $\rhohat(\vb*{x})$ and a set of system observables $\mathcal{O}=\{ \hat{O}_j\,|\, j = 1,\dots, P \}$, information about the response of the open quantum system to the input $\vb*{x}$ is contained in the following vector
\bea
    \vb*{\phi}(\vb*{x}) \equiv \(1, \langle \hat{O}_1\rangle_{\vb*{x}}, \dots,\langle \hat{O}_P\rangle_{\vb*{x}} \)^T,
\label{eq:generalized_Bloch_vector}
\eea
where 
\bea
\langle \hat{O}_j \rangle_{\vb*{x}} = \tr[\hat{O}_j\rhohat(\vb*{x})].
\eea

The vector $\vb*{\phi}(\vb*{x})$ belongs to the feature space $\mathcal{F}\subseteq\mathbb{R}^P$ and depends on the input $\vb*{x}$, generally in a nonlinear fashion. Note that the constant component $1$, which ensures that the trial function can fit a biased target function, can be seen as the measurement of the identity observable, since the density matrix $\rhohat(\vb*{x})$ always has unit trace.

Finally, the trial function $f$ of the noisy quantum kernel machine, which depends on the vector of variational parameters $\vb*{w}$, is given by the affine transformation:
\bea
   f:\vb*{x}\mapsto \vb*{w}^T\vb*{\phi}(\vb*{x}),
\eea
where the vector $\vb*{w} \equiv \(b, w_1, \dots, w_P\)^T\in\mathbb{R}^{P+1}$ contains the parameters of the linear transformation and $b$ represents the bias term.  
An alternative approach to the construction of the feature vector, based on time-multiplexing measurements, will be presented in Sec. \ref{section:driven_dissipative_spins_model}.

\subsection{Training procedure}
\label{section:training_procedure}

A trial function characterized by its weights $\vb*{w}$ can be optimized using a regularized least-squares loss function over a training set $(\vb*{x}_i,{y}_i) \in \mathcal{S}$ consisting of $\ntrain$ inputs $\vb*{x}_i \in \mathcal{X}$ and labels ${y}_i \in  \mathcal{Y}$, namely:
\bea
\mathcal{L}\left(\vb*{w} \mid \mathcal{S}\right) :=\frac{1}{2\ntrain}\sum_{i=1}^{\ntrain}\(y_i - \vb*{w}^T\vb*{\phi}(x_i)\)^2
+\frac{\lambda}{2}\norm{\vb*{w}}_2^2 \, .
\label{eq:L2loss_with_reg}
\eea
The second term in Eq.~\eqref{eq:L2loss_with_reg} is a regularization penalty that helps to prevent overfitting. The corresponding regularization parameter $\lambda$ controls the strength of the overfitting penalty. Adding such a regularization bias is on average equivalent to adding a centered Gaussian noise of variance $\lambda$ to the measurement features before the optimization~\cite{goodfellow2016}.

Such classifiers are known as least-square support-vector classifiers~\cite{suykens1999}. Although most classification problems are commonly treated with other loss functions~\cite{hastie2013}, using the least-squares loss function allows us to perform the optimization analytically. Indeed, upon introducing the $(P+1)\times \ntrain$ matrix $\vb{\Phi}$, whose columns are the quantum feature vectors $\vb*{\phi}(\vb*{x}_i)$ associated to the training input $\vb*{x}_i$, and $\vb*{y}$, the column vector of size $\ntrain$ containing the corresponding labels, the optimal weights are given by \footnote{Note that in certain scenario, one may prefer to exclude the first component of $\vec{w}$, which is a constant intercept term, in the regularization. Then it suffices to replace the $(1,1)$ entry of $\id$ by $0$ in Eq.~\eqref{eq:optimal_weights} to obtain the optimal weights. This is equivalent to using a centered kernel, as discussed in Appendix \ref{appendix:B_intercept_and_kernel_centering}.}
\bea
\vb*{w}^{\ast} = \(\vb{\Phi}\vb{\Phi}^T + \ntrain\lambda\id\)^{-1}\vb{\Phi}\vb*{y} \,.
\label{eq:optimal_weights}
\eea

\section{Quantum kernel and decoherence}
\label{section:intro_quantum_kernel_and_dissipation}

The generic encoding-decoding scheme encompasses a large class of quantum machine-learning models. Here, we describe a decoding based on a linear combination of measurements, but other decoding methods were proposed in the literature. In particular, it was recently shown that quantum neural networks can be mapped to models with an encoding/decoding structure~\cite{jerbi2021}, where the decoding is achieved by optimizing a single parametrized measurement.

Models described by the previous scheme can be analyzed in the framework of kernel theory, which provides useful tools to understand properties such as expressivity, trainability and capacity to generalize to a test sample of unseen data. In this section we first concisely introduce the kernel framework. We then specialize our discussion to noisy quantum kernels, and show how we can link the role of dissipation and decoherence to the kernel's main figures of merit.

We aim at determining the largest class of functions that can be approximated by our trial function $f$. This class depends on the type of decoding used, that is on the specific set of measurements that are performed on the quantum system. When measuring a set ${\mathcal O}$ of observables, this function space reads
\bea
    \mathcal{H}({\mathcal O}) = \{f:\vb*{x} \mapsto \tr [\rhohat(\vb*{x})\hat{A}]\,\mid\, \hat{A}\in \mathrm{Span}({\mathcal O})\} \, .
\eea
In this case, the feature vector $\vb*{\phi}(\vb*{x})$ gives rise to a positive semi-definite and symmetric function which we call the feature kernel:
\bea
k_{\mathcal{O}}(\vb*{x},\vb*{x}') = \vb*{\phi}(\vb*{x})^T\vb*{\phi}(\vb*{x'}).
\label{eq:feature_kernel}
\eea
This kernel function, together with the probability distribution $p$ of inputs $\vb*{x}\in\mathcal{X}$ \footnote{The probability measure $p$ on the input space $\mathcal{X}$ is important here as it determines the scalar product on the space of real-valued functions on $\mathcal{X}$ through {$\langle f,g\rangle = \expectt{\vb*{x}\sim p}{f(\vb*{x})g(\vb*{x})}$}. The reproducing property, crucial to link the RKHS and its kernel, relies on such a well-defined scalar product.}, uniquely determines a specific set of real-valued functions, the so-called reproducing kernel Hilbert space (RKHS):
\bea
 \mathrm{Span}\{f:\vb*{x}\mapsto k_{\mathcal{O}}(\vb*{x},\vb*{x}')\,\mid\,\vb*{x}'\in\mathcal{X}\}.
\eea
The RKHS associated to $k_{\mathcal{O}}$ can be shown to be exactly the space of hypothesis functions $\mathcal{H}(\mathcal{O})$~\cite{paulsen2016}. Hence, the study of the kernel function allows one to investigate the structure of $\mathcal{H}(\mathcal{O})$. In particular, it follows that one can use the eigendecomposition of the kernel function as a basis of the class of functions that can be represented by our model. This useful property motivates the adoption of a kernel standpoint in what follows.  

\subsection{Quantum kernel}

In order to discuss the expressive power of our model, we introduce the largest class of transformations $\mathcal{H}_{\text{full}}$ that can be achieved for a given encoding strategy~\cite{schuld2021a}:
\bea
    \mathcal{H}_{\text{full}} = \{f:\vb*{x} \mapsto \tr [\rhohat(\vb*{x})\hat{A}]\,\mid\,\hat{A} = \hat{A}^{\dagger} \} \, .
\eea
The class of transformation yielded by a set of measurements $\mathcal{O}$ is necessarily included in this maximal class $ \mathcal{H}({\mathcal O}) \subseteq \mathcal{H}_{\text{full}}$; the equality holds whenever $\mathcal{O}$ is a complete set of observables.
In the following, we will use the term ``full tomography" to refer to this ideal implementation.
It turns out that $\mathcal{H}_{\text{full}}$ is the RKHS of a particular kernel, the quantum kernel, that solely depends on the feature map $\rhohat(\vb*{x})$~\footnote{For a closed system, this quantum kernel can be directly evaluated through measurement \cite{park2020} and the trial function can be expressed in terms of the quantum kernel and optimized in an equivalent way.} \cite{schuld2021}:
\bea\label{eq:quantum_kernel_function}
    k(\vb*{x},\vb*{x}') = \tr\[\rhohat(\vb*{x})\rhohat(\vb*{x}')\].
\eea
This kernel arises naturally from the Hilbertian structure of the space of quantum states. As it represents the maximal achievable class of transformation an encoding can give, the quantum kernel provides insight on the expressive power of our model. Note that this kernel can be identified with the previous feature kernel $k_{\mathcal{O}}$ provided that the measurements $\mathcal{O}$ form an orthonormal basis $\mathcal{B} = \{B_j\}_j$ of the space of observables, i.e. $k = k_\mathcal{B}$ with $\tr[\hat{B}_i\hat{B_j}]=\delta_{ij}$ and we impose $\hat{B}_0 \propto \hat{\id}$ by convention.

In what follows, it will be useful to work with a ``centered" version of the quantum kernel. Centering the kernel is equivalent to working with hypothesis functions that have zero mean value on the input set. As we will show, this is convenient for interpreting some of the key quantities we will introduce in terms of probabilistic quantities.
In Appendix \ref{appendix:B_intercept_and_kernel_centering}, we show that, at least for balanced data, the use of the L2 loss function allows us to work with a centered version of the quantum kernel without lack of generality. The centered kernel is given by
\bea
k_c(\vb*{x},\vb*{x}') &= \tr\[\delta\rhohat(\vb*{x})\delta\rhohat(\vb*{x}')\]
\eea
and the corresponding RKHS is
\bea
\mathcal{H}_{k,c} = \mathrm{Span}\{f:\vb*{x}\mapsto k_c(\vb*{x},\vb*{x}')\mid \vb*{x}'\in\mathcal{X}\}.
\eea
The constant feature we introduced in Eq.~\eqref{eq:generalized_Bloch_vector} becomes irrelevant when using centered quantities, so we drop it and define
\bea
\delta \vb*{\phi}(\vb*{x}) \equiv \left(\delta\langle \hat{O}_1\rangle_{\vb*{x}}, \dots,\delta\langle \hat{O}_P\rangle_{\vb*{x}} \right)^T.
\eea
We can also correspondingly drop the weight term $b$, so that the weight vectors can be redefined as $\vb*{w} = (w_1,\dots,w_P)^T\in\mathbb{R}^P$. The space $\mathcal{H}_{k,c}$ can be rewritten as
\bea
\mathcal{H}_{k,c} = \{f:\vb*{x} \mapsto \vb*{w}^T\delta\vb*{\phi}(\vb*{x}),\quad \vb*{w}\in\mathbb{R}^{P} \},
\eea
where the centered quantum kernel reads
\bea
k_c(\vb*{x},\vb*{x}') = \delta\vb*{\phi}(\vb*{x})^T\delta\vb*{\phi}(\vb*{x}') \, 
\eea
with the choice of $\mathcal{O} = \mathcal{B}$.
The quantum feature matrix $\vb{\Phi}$ is then replaced by a $P\times \ntrain$ matrix $\delta\vb{\Phi}$, whose columns are the centered feature vectors $\delta\vb*{\phi}(\vb*{x_i})$.

\subsection{Kernel eigen-decomposition}

Under general assumptions, the centered quantum kernel admits a decomposition into an orthonormal family of eigenfunctions \cite{paulsen2016}: 
\bea
    k_c(\vb*{x},\vb*{x}') &= \sum_i \lambda_i \delta\psi_i(\vb*{x})\delta\psi_i(\vb*{x}')\, ,\\
    \expectt{p}{\delta\psi_i\delta\psi_j} &= \delta_{ij}\, ,
\eea
where $\lbrace\lambda_i\rbrace_i$ are positive eigenvalues sorted in a decreasing order, namely $\lambda_{i+1}\leq\lambda_{i}$, $\forall i$. When necessary, we can complete this orthonormal family into a basis with eigenfunctions associated to zero eigenvalues. In the case of the uncentered quantum kernel, the kernel eigenfunctions correspond to an orthonormal basis of system observables~\cite{kubler2021}. When the kernel is centered, the basis of kernel eigenfunctions corresponds to an orthonormal basis $\{\hat{E}_i\}_i$ of the space of zero-trace observables, which we call eigenobservables. Such operators satisfy the following properties:
\bea
\tr\[\hat{E}_i\hat{E}_j\] &= \delta_{ij}\, ,\\
\tr\[\hat{E}_i\] &= 0\, .
\eea
The eigenfunctions are given by
\bea
\delta\psi_i(\vb*{x}) &= \frac{1}{\sqrt{\lambda_i}}\tr\left[\delta\rhohat(\vb*{x})\hat{E}_i\right]\\
&= \frac{1}{\sqrt{\lambda_i}}\delta\langle\hat{E}_i\rangle_{\vb*{x}} \, .
\eea
The corresponding eigenvalues are then given by the variances of the eigenobservable measurements over the input set, namely:
\bea
    \lambda_i = \expectt{p}{\delta\langle\hat{E}_i\rangle_{\vb*{x}}^2}
    = \mathrm{Var}_p\[\langle\hat{E}_i\rangle_{\vb*{x}}\] \,.
\eea
One can see this eigen-decomposition of the kernel as a principal-component analysis in the space of quantum features, as it yields an orthogonal basis of measurement functions ordered by their variances on the input set. We stress that these are variances of the observables expectation values over the quantum states representing the different inputs, and thus are very different from the quantum variance of the corresponding observable for a specific state.

The previous decomposition of the kernel is very useful for grasping the learning mechanism and the model expressivity. Upon working with centered features, the loss function introduced in Eq.~\eqref{eq:L2loss_with_reg} becomes
\bea
\mathcal{L}_c\left(\vb*{w} \mid \mathcal{S}\right) =\frac{1}{2\ntrain}\sum_{i=1}^{\ntrain}\(y_i - \vb*{w}^T\delta\vb*{\phi}(x_i)\)^2
+\frac{\lambda}{2}\norm{\vb*{w}}_2^2 \,.
\label{eq:L2loss_with_reg_centered}
\eea
Following~\cite{hastie2013}, we can decompose the trial function $f(\vb*{x}) = \vb*{w}^T\delta\vb*{\phi}(\vb*{x})$ in the basis of the kernel eigenfunctions, namely as $f(\vb*{x}) = \sum_j\beta_j\delta\psi_j(\vb*{x})$. Exploiting such decomposition, the loss function becomes
\bea
\mathcal{L}_c\left(\vb*{\beta} \mid \mathcal{S}\right) =&\frac{1}{2\ntrain}\sum_{i=1}^{\ntrain}\Bigl[y_i - \sum_j\beta_j\delta\psi_j(\vb*{x}_i)\Bigr]^2\\
&+\frac{\lambda}{2}\sum_j\frac{\beta_j^2}{\lambda_j} \, .
\eea
Note that in the regularization term the components of the trial function on the eigenbasis are weighted by the corresponding kernel eigenvalues. The lower the variance of an eigenobservable, the more the corresponding eigenfunction is penalized. Hence the regularization parameter $\lambda$ acts as a smooth cutoff on the basis of the kernel eigenfunctions, which are then used to approximate the target function.

The spectrum of the kernel characterizes the generalization capacity and the expressivity of our model. It also finds applications in understanding many other machine learning scenarios. For instance, in the context of classical neural networks it has links with learning curves~\cite{bordelon2021, canatar2021}. Moreover, the kernel (or the neural tangent kernel in the context of classical neural networks) shares its spectrum with the Fisher information matrix, of particular relevance for quantum neural networks \cite{abbas2021}.

\subsection{Role of decoherence on expressivity and generalization error}
\label{section:expressivity_generalization_and_dissipation}
The exponential growth of the Hilbert space dimension with the number of qubits in a network and the complex dynamics of quantum systems have created hope for a quantum advantage in the field of quantum machine learning. However, it is known that having a very high-dimensional feature space does not necessarily guarantee high machine-learning performances~\cite{hastie2006, hastie2013}. Indeed, recent investigations within the quantum kernel framework somehow mitigated the hope for a general quantum advantage \cite{huang2021, wang2021, kubler2021}. Yet, a clear quantum advantage has been demonstrated for some specific tasks \cite{liu2021, wu2021}, again by exploiting the quantum-kernel formalism. 
In order for a quantum-kernel-based model to perform well on a given task, the set of transformations achieved must be well ``aligned" with the target function $y(\vb*{x})$. This notion of alignment is mathematically encapsulated in the kernel-target-alignment measure \cite{cristianini2006} which reads, for the centered quantum kernel,
\bea\label{eq:alignment}
    A(k_c,y) &= \dfrac{\expectt{p}{y(\vb*{x})k_c(\vb*{x},\vb*{x}')y(\vb*{x}')}}{\expectt{p}{k_c(\vb*{x},\vb*{x}')^2}^{1/2}\expectt{p}{y(\vb*{x})^2}}\\
    &= \dfrac{\sum_i \lambda_i \expectt{p}{\delta\psi_i(\vb*{x})y(\vb*{x})}^2 }{(\sum_i \lambda_i^2)^{1/2}\expectt{p}{y(\vb*{x})^2}} \, .
\eea

Although the kernel-target alignment measures how well a kernel and the associated embedding fits a specific function, in this article we introduce another figure of merit that does not depend on a specific task, namely the ``effective kernel rank" $R_{\rm{eff}}(k)$, which quantifies the effective number of independent transformations that a given kernel can yield. Such a quantity is defined as:
\bea
    \sqrt{R_{\rm{eff}}(k_c)}=&\sum_{j}A(k_c,g_j) \, ,
\eea
where $\{g_j\}_j$ is any orthonormal basis of functions on the input space. As shown in Appendix \ref{appendix:A1_kernel_expressivity_and_effective_rank}, 
for the centered quantum kernel, the effective kernel rank can be also expressed in terms of variances of the quantum expectation values of the measured observables:
\bea
    \sqrt{R_{\rm{eff}}(k_c)} =\frac{\sum_{i=1}^P \mathrm{Var}_p\[\meanq{\hat{O}_i}_{\vb*{x}}\]}{\(\sum_{i,j=1}^P\mathrm{Cov}_p\[\meanq{\hat{O}_i}_{\vb*{x}},\meanq{\hat{O}_j}_{\vb*{x}}\]^2\)^{\frac{1}{2}}} \, .
\eea
Note that the denominator acts as a normalization and can be seen as a measure of the redundancy of the embedding when expressed in terms of $\hat{O}_i$. 
In section  \ref{section_numerical_results}, we will investigate in a rather general class of physical models how the kernel effective rank scales with the system size and with noise.

In Appendix \ref{appendix:A1_kernel_expressivity_and_effective_rank}, we also provide the proof showing that the kernel effective rank can be expressed in terms of the kernel spectrum:
\bea
    \sqrt{R_{\rm{eff}}(k)} = \frac{\sum_i \lambda_i}{\sqrt{\sum_i \lambda_i^2}}\, .
    \label{eq:kernel_quality}
\eea
This expression is reminiscent of the reciprocal of the inverse participation ratio. The kernel effective rank provides information about the size of its support. Moreover, we have the following inequality:
\bea
    R_{\rm{eff}}(k)\leq \vert{}\{ \lambda_i\neq 0 \}\vert{} \, .
\eea
This is saturated when all the non-zero eigenvalues are all equal. The numerator in the expression for the square-root of the effective kernel rank is the kernel trace, which can be rewritten as:
\bea
    \sum_i \lambda_i = \expectt{p}{\tr\[\rhohat(\vb*{x})^2\]}-\tr\[\expectt{p}{\rhohat(\vb*{x})}^2\] \,.
    \label{eq:kernel_trace_and_purity}
\eea
In this expression, we recognize the difference between the average purity of the embedded density matrices over the input space and the purity of the average embedding matrix. The first term is of great relevance to our study, as it crucially depends on the dissipation and decoherence affecting the noisy quantum system: indeed, a low purity is the consequence of the openness of the quantum system.
The second term instead measures the diversity of the embedding map; its importance is discussed in \cite{kubler2021}.

We emphasize that the kernel trace also appears to be relevant when investigating the ability of the model to perform well on unseen data, hence on its generalization properties. To measure the performance of a model on a binary classification task we use the accuracy $\mathcal{A}$. Given a prediction function $f$, the accuracy is given by the fraction of samples for which $f$ assigns the right label and it can be defined as the expectation of a 0-1 loss function:
\bea
\mathcal{A}(f) &= \expect{\id_{y(\vb*{x})f(\vb*{x})\geq0}} \, .
\eea
Since during the training we only have access to the data set $\mathcal{S}$ and not to the true distribution $p$, expectations values can only be approximated using the empirical distribution $\hat{p}$ on $\mathcal{S}$. The corresponding empirical expectations are given by $\expectt{\hat{p}}{f(\vb*{x})} = \frac{1}{\ntrain}\sum_{i=1}^{\ntrain} f(\vb*{x}_i)$. From this, we can define the empirical accuracy $\mathcal{A}$ on the training set $\mathcal{S}$ and the true accuracy $\mathcal{A}^{*}$. Correspondingly, we can introduce the risk $\mathcal{R}^*$, also called error or inaccuracy, as $\mathcal{R}^* = 1 - \mathcal{A}^*$ (its empirical counterpart is defined analogously). It is convenient to introduce  slightly modified versions of the  risk and inaccuracy that depend on a margin-parameter $\eta > 0$. We introduce the $\eta$-margin loss as:
\bea
    \Phi_{\eta}(y) = \begin{cases}
      1 & \text{if}\quad y\leq0\\
      1-\frac{y}{\eta} & \text{if}\quad 0\leq y\leq \eta\\
      0 & \text{if}\quad \eta\leq y 
    \end{cases} \,.
\eea
Correspondingly, we can introduce the empirical $\eta$-margin risk as:
\bea
    \mathcal{R}_{\eta}(f) = \expectt{\hat{p}}{\Phi_{\eta}(y(\vb*{x})f(\vb*{x}))}\, .
\eea
The $\eta$-margin-risk and the risk satisfy the following inequality:
\bea
\mathcal{R}(f) \leq \mathcal{R}_{\eta}(f)\leq \expectt{\hat{p}}{\id_{y(\vb*{x})f(\vb*{x})\leq\eta}}\, .
\eea

The ability of the model to generalize well on unseen data is then quantified by the generalization error:
\bea
\mathcal{E} = \mathcal{R}^* - \mathcal{R} \,.
\eea
For kernel methods with kernel $k$, the generalization error admits an upper-bound involving the $\ntrain \times \ntrain$ empirical kernel matrix $\vb{K}$ whose entries are defined as $K_{ij} = k(\vb*{x}_i,\vb*{x}_j)$. This bound depends on the specific task under consideration and on the exact space of trial functions used (details on the bound used and its derivation can be found in \cite{mohri2018foundations} and in Appendix \ref{appendix:A2_generalization_bound_kernel}). 

To derive the upper-bound, we fix a class of trial functions of the form $f: \vb*{x} \mapsto \vb*{w}^T\delta\vb*{\phi}(\vb*{x})$ where $\delta\vb*{\phi}(\vb*{x})$ corresponds to measurements of an orthonormal basis of observable: $\delta\phi_i(\vb*{x}) =\delta \meanq{\hat{B}_i}_{\vb*{x}}$. We further constrain this class by choosing a parameter $\Lambda \geq 0$, and require that the trial function's parameters $\vb*{w}$ satisfy $\norm{\vb*{w}}^2\Lambda\leq 1$. By exploiting Eq.~\eqref{eq:kernel_trace_and_purity}, we get that, for such functions, the following inequality holds with probability at least $1-\delta$ on the training set $\mathcal{S}$:
\bea
\mathcal{R}^*(f)- \mathcal{R}_{\eta}(f) \leq&\frac{2}{\eta}\(\frac{\expectt{\hat{p}}{\tr\[\rhohat^2\]} - \tr\[\expectt{\hat{p}}{\rhohat}^2\]}{\ntrain\Lambda}\)^{\frac{1}{2}}\\
&+3\sqrt{\frac{\mathrm{log}(\frac{2}{\delta})}{2\ntrain}}\, .
\label{eq:generalization_error_and_dissipation}
\eea
Other generalization bounds can be established, in particular the authors of \cite{banchi2021} found another bound using a quantum information theory standpoint, and their conclusions are in agreement with our results.
Let us make a few important comments on the meaning of this inequality. 
The inequality has a probabilistic character controlled by $\delta>0$. If we set this parameter to $0^{+}$, the bound is always satisfied although it becomes trivial. The same goes with the margin parameter $\eta$: as $\eta\to 0^{+}$ the margin-error $\mathcal{R}_{\eta}(f)$ tends to the training error $\mathcal{R}(f)$, but again the right-hand side of the inequality diverges. The parameter $\Lambda$ is another sort of regularization parameter, as the parameter $\lambda$: if $\Lambda \to 0^{+}$, the norm of the weight vector $\norm{\vb*{w}}$ can be arbitrarily large and overfitting is not limited. Correspondingly, the right-hand side diverges and the bound becomes trivial.
The most important crucial physical quantity involved in the upper bound is the kernel trace given in Eq.~\eqref{eq:kernel_trace_and_purity}. Such a quantity accounts for the model expressivity. This duality between expressivity and generalization is crucial in machine learning \cite{hastie2013}. What is relevant to our study is that this expressivity measure involves the mean purity of the embedded states and hence is affected by dissipation and decoherence acting on the noisy quantum kernel machine. The appearance of the regularization parameter $\Lambda$ in this upper bound is also relevant as it allows us to establish a link with experimental constraints, such as imperfect measurements. In fact, as we will see in Section \ref{section_numerical_results}, adding a Gaussian error of standard deviation $\sigma$ to the observable measurements is equivalent to working with an infinitely precise measurement apparatus while replacing the regularization parameter $\lambda$ with $\lambda + \sigma^2$ \cite{goodfellow2016}.

\section{Noisy quantum kernel machines with driven-dissipative spin chains}
\label{section:driven_dissipative_spins_model}

As an illustrative example, we here numerically simulate noisy quantum kernel machines based on 1D chains of spins subject to both driving and decoherence.

\begin{figure}[t!]
    \centering
    \includegraphics[width=0.5\textwidth]{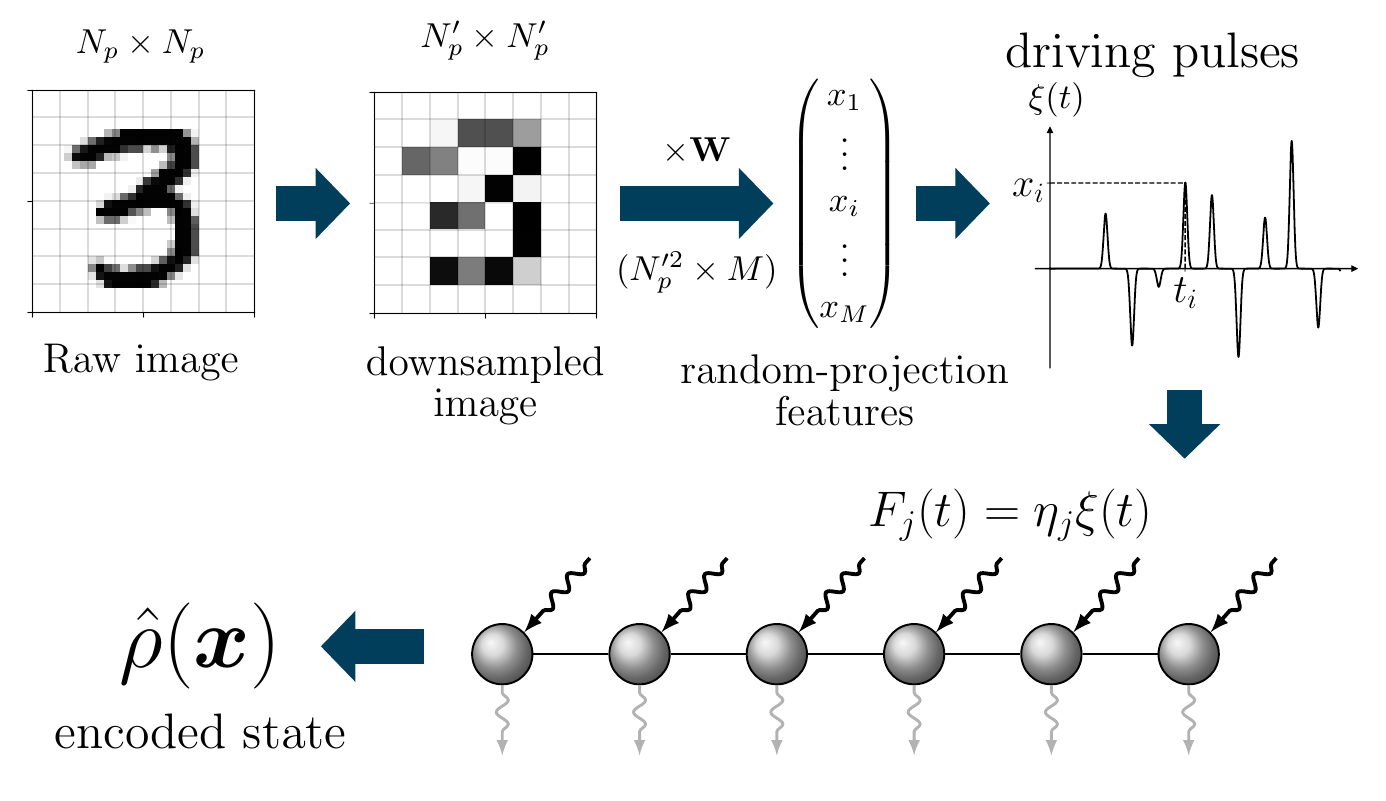}
    \caption{Schematic representation of the encoding procedure for the MNIST classification task. The input grayscale image, of original size $N_p\times N_p$, with $N_p = 28$, is first downsampled to a size of $N_p'\times N_p'$ (or $N_p'^{2}\times 1$ when viewed as a column vector), with $N_p'=8$, and linearly transformed by a fixed $N_p'^2\times M$ random projection filter $\vb{W}$ to yield the vectors $\vb*{x'}$ containing $M=10$ random-projection features. Those features are normalized by 3 times the standard deviation over the set of all features for all images in the training set, and we denote $\vb*{x}$ the normalized vectors representing the images. The vector $\vb*{x}$ is then encoded into a sequence of driving pulses $\xi(t)$, where the amplitude of the $i$th pulse (at time $t_i$) is proportional to the input's $i$th component $x_i$. Finally, the pulses are used to drive a spin chain (initially prepared in the state $\rhohat_0$), where the driving amplitude at site $j$ is $F_j(t) = \eta_j\xi(t)$ with $\eta_j$ a random site-dependent scale factor. We define the state of the spin chain immediately after the driving sequence to be the encoded state, represented by its density matrix $\rhohat(\vb*{x})$.}
    \label{fig:enc_mnist}
\end{figure}

The simulation of such an open quantum system for a large number of inputs, various choices of the number of sites and distinct disorder realizations is a computationally daunting task~\footnote{In total, the results presented in this work required approximately $800{,}000$ scalar hours ($\sim 90$ years) of computation and $5$ terabytes of storage on the acknowledged French National High Performance Computing facility (GENCI). During the simulations, we used approximately up to $30,000$ cores simultaneously.}. Indeed, this requires to exactly integrate a large set of corresponding Lindblad master equations of the form of Eq.~\eqref{eq:mastereq}. Hence, we have considered a simplified classification task involving only a subset of the MNIST dataset, namely classifying images of handwritten digits corresponding to the digits $3$, $6$ and $8$, which share common shapes. 
A schematic description of the task and of the feature encoding through driving of the considered physical system is presented in Fig.~\ref{fig:enc_mnist}. 

\begin{figure}[t!]
    \centering
    \includegraphics[width=0.5\textwidth]{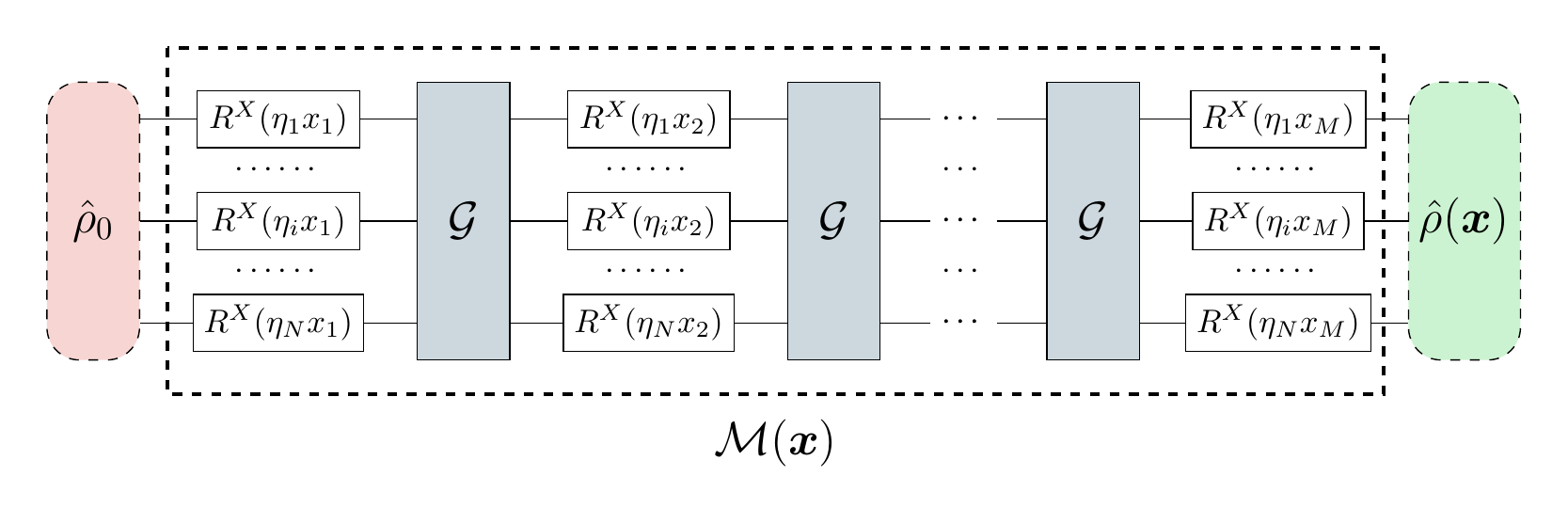}
    \caption{Equivalent circuit of the encoding procedure for the MNIST classification task. If the driving pulses are sharp enough, the encoding process of Fig. \ref{fig:enc_mnist} can be equivalently seen as a quantum circuit, where the $i$-th driving pulse on site $j$ is effectively a single-qubit $X$-rotation gate $R^X(\eta_j x_i)$, and the pulses at different times are separated by the gate $\mathcal{G}$ generated by the free dynamics of the spin system in the absence of the drive. Note that the entire process between $\rhohat_0$ and $\rhohat(x)$ serves as the dynamical map $\mathcal{M}(x)$ shown in Fig. \ref{fig:dec_sch}.}
    \label{fig:circuit}
\end{figure}

The original MNIST dataset consists of $28\times28$-pixel images. Encoding such high-dimensional features in the state of a quantum system is not an easy task. Therefore, we first linearly down-sample the raw images from $28\times28$ to $8\times8$ pixels, thereby reducing the dimension of the input features. The down-sampled images, viewed as vectors, are then multiplied by a random $8^2\times10$ matrix $\vb{W}$, whose entries are uniformly drawn over the interval $[-1, 1]$, yielding vectors $\vb*{x'}=(x'_1,\dots, x'_{M})^T$ of $M=10$ random-projection features. These are finally normalized by 3 times the standard deviation of the set $\{x'_i \mid i=1,\dots,M,~x\in\mathcal{S}\}$. At the end of this procedure, every image in the dataset is represented by a vector $\vb*{x}$ of size $M=10$, which will be used as inputs in the following. These are computed only once and reused throughout this article, except in section~\ref{section:optimized}.

\begin{figure}[tp!]
    \includegraphics{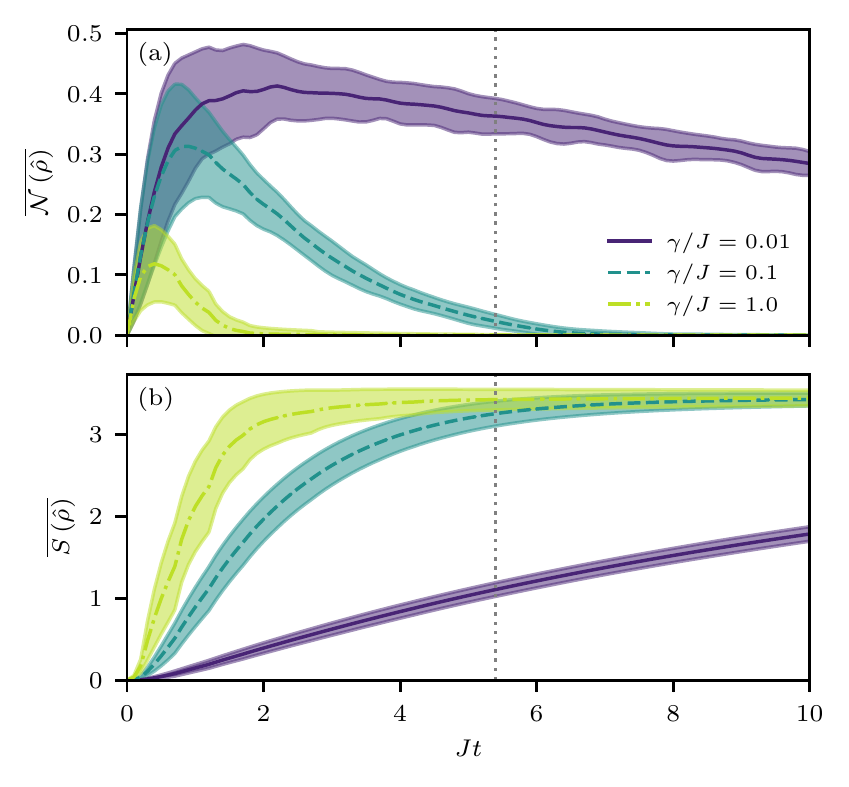}
    \caption{Time dynamics of the average entanglement negativity $\overline{\mathcal{N}(\rhohat)}$ (a) and the average von Neumann entropy $\overline{S(\rhohat)}$(b) in presence of pure dephasing with different values of the corresponding rates $\gamma$. At the initial time $t=0$ the system is in the pure state $\hat{\rho}_0$ defined in the text. Note that the driving sequence finishes at the time $J t \simeq 5$ indicated by the vertical dotted lines on the figures. The time is expressed in units of $1/J$ where $J$ is the average value of the spin coupling. We define $\overline{\mathcal{N}(\rhohat)}$ as the average over all the sites of the negativities associated to the system partitions having the form $\{\{\mathrm{site} \; i\},\{\mathrm{site}\; j \mid j\neq i\}\}$.  This quantity is averaged over $20$ inputs $\vb*{x} \in \mathcal{X}$, $5$ disordered configurations of spin couplings and for a chain of $N=5$ spins. The filled areas correspond to a one standard deviation confidence interval. }
    \label{fig:encoding_dynamics}
    \phantomlabel{a}{fig:entanglement_vs_time}
    \phantomlabel{b}{fig:entropy_vs_time}
\end{figure}

\begin{figure*}[ht!]
    \includegraphics{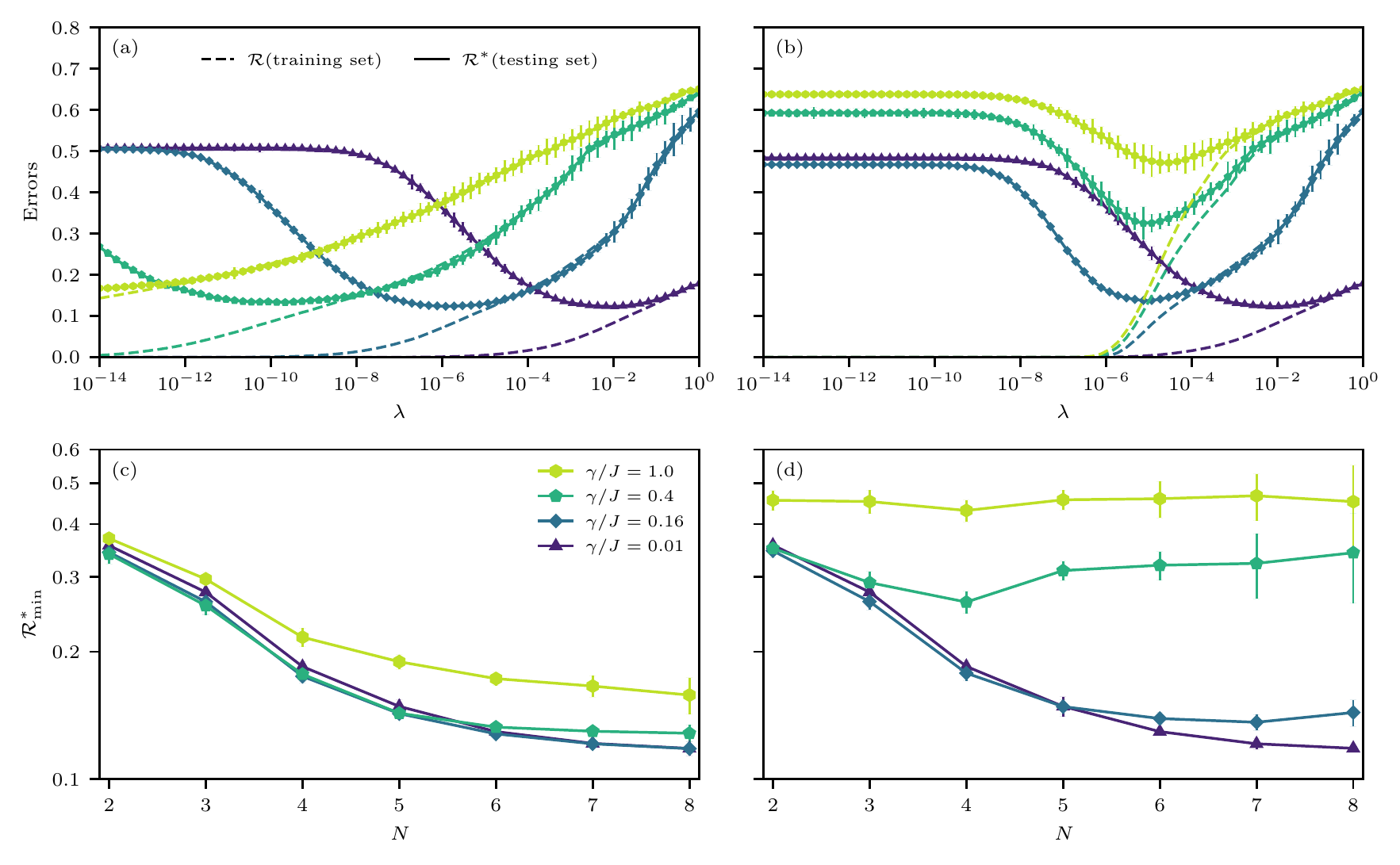}
    \caption{(a) Training error $\mathcal R$ (dashed lines) and testing error $\mathcal R^*$ (solid lines and markers) as a function of the regularization parameter $\lambda$ for a chain with $N=7$ spins in the presence of pure dephasing for different values of the corresponding rate $\gamma$ (different markers in the legend) in units of the average spin coupling $J$. Measurements are assumed to be ideal . (b) Same as (a) with an extra random Gaussian noise of width $\sigma=10^{-3}$ added to the observable expectation values to account for imperfect measurements. For each value of $\lambda$ the corresponding errors are averaged over $15$ disordered configurations, and the error bars are bootstrap estimates of the standard deviation for the estimated mean values. We use $10$ bootstrap sets, each consisting of $15$ samples randomly drawn with replacement from the original set of $15$ disorder realizations.
    (c) Minimal testing error as a function of the number of spins $N$ for different values of the dephasing rate $\gamma$. For each disorder configuration, the regularization parameter $\lambda$ is chosen to minimize the testing error and the resulting minimum is averaged over the disorder. The error bars are derived using the same bootstrap procedure. Number of disorder configurations: $50$ for $N=2$ to $N=5$ spins, $25$ for $N=6$, $15$ for $N=7$ and $5$ for $N=8$. (d) Same as (c) with an extra random Gaussian noise of width $\sigma=10^{-3}$ added to the observable expectation values to account for imperfect measurements.}
    \label{fig:full_tomo_results}
    \phantomlabel{a}{fig:full_tomo_results:a}
    \phantomlabel{b}{fig:full_tomo_results:b}
    \phantomlabel{c}{fig:full_tomo_results:c}
    \phantomlabel{d}{fig:full_tomo_results:d}
\end{figure*}

This encoding is designed so as to fix the amount of information fed to the system, independently from its number of sites. It allows us to perform a fair comparison of models associated to quantum systems of increasing sizes. In particular, this ensures that any observed increase of the performance with the system size is solely due to an intrinsic enhancement of the model expressive power. In Section~\ref{section:optimized}, we lift the above-defined ``information bottleneck'' and use a different encoding, where the number of encoded features $M$ scales with the system size $N$. Therein, we show that this results in competitive performances, as compared to classical reservoir-computing settings involving hundreds to thousands of degrees of freedom~\cite{opala2019, ballarini2020}.

In what follows, we denote $\mathcal{X} \subseteq \mathbb{R}^{M}$ the input space consisting of the random-projection features representing the images to classify, and $\mathcal{Y} = \{3,6,8\}$ the set of corresponding labels. Our dataset consists of 17{,}000 images, which we split into a training set of $\ntrain =15{,}000$ images and a testing set of $\ntest=2000$ images. As before, the training set is denoted as $\mathcal{S} = \{(\vb*{x}_i, y_i)\in\mathcal{X}\times\mathcal{Y}\,|\, i=1,\dots, \ntrain\}$.

The system in which we encode the previous features is a driven-dissipative one-dimensional chain of $N$ spins-$1/2$ described by the following Heisenberg XYZ Hamiltonian:
\bea
    \Hhat(t; \vb*{x}) = &\frac{\hbar}{2}\sum_{i=1}^N\left(F_i(t; \vb*{x})\hsx^i+\Delta_i\hsz^i\right)\\
    -& \frac{\hbar}{2}\sum_{\langle i,j \rangle} (J^x_{ij}\hsx^i\hsx^j+J^y_{ij}\hsy^i\hsy^j+J^z_{ij}\hsz^i\hsz^j),
\label{eq:Hamiltonian_XYZ}
\eea
with $F_i(t; \vb*{x})$ an input-dependent driving field, $\Delta_i$ an on-site frequency detuning, and $J^{k}_{ij}$ the symmetric coupling rate between nearest neighbors. Here, indices $\langle i,j \rangle$ run over all pairs of nearest neighbors. Parameters $J^{k}_{ji}$ and $\Delta_i$ are uniformly drawn at random in the interval $[0, 2J]$. $1/J$ will be used as unit of time in the numerical plots. We prepare the system in an initial state with all spins down $\rhohat_0 = \bigotimes_{i=1}^N\ketbra{0}$.

The encoding of the input $\vb*{x}$ corresponding to a given image into the system state is performed by driving the system with a series of $M=10$ sharp Gaussian pulses, whose amplitudes are proportional to the input vector elements, as illustrated in Fig.~\ref{fig:enc_mnist}.
We first define a generic driving $\xi(t;\vb*{x})$ from the feature $\vb*{x}$:
\bea
\xi(t;\vb*{x}) &= \sum_{k=1}^N \frac{x_{k}}{\sqrt{2\pi}\sigma}\exp\(-\frac{(t-t_k)^2}{2\sigma^2}\) \, ,\\
    t_k &= (k-1)\Delta t+10\sigma, \quad\forall k=1,\dots,M \, ,
\label{eq:generic_driving}
\eea
where the time interval between two successive pulses is $\Delta t = 1/(2J)$ and the width of each pulse is $\sigma =1/(50J)$. Then the driving on site $i$ is taken to be proportional to this generic driving:
\bea\label{eq:on-site_driving}
    F_i(t;\vb*{x}) = \eta_i  \xi(t;\vb*{x}),
\eea
where the $\eta_i$ are random factors uniformly distributed in the interval $[-\pi, \pi]$. Under these driving conditions, the coherent part of the system dynamics can be thought of as that of an equivalent quantum circuit alternating between a set of local $X$-rotation gates, of the form $R_i^{X}(\eta_i x_k)$, and a deep block generating entanglement among qubits~\footnote{This can be explicitly expressed as a $D$-deep circuit via Trotterization as $\mathcal{G} = \[\prod_{i=1}^N R_i^Z\(-\tfrac{2\Delta_i \Delta t}{D}\) \prod_{\langle i, j \rangle}\prod_{K\in\{X,Y,Z\}} R_{ij}^{KK}(\tfrac{2 J_{ij}^K \Delta t}{D}) \]^D + O(J_0^2 \Delta t^2/D^2)$}, as illustrated in Fig.~\ref{fig:circuit}. The scaling factors $\eta_i$ prevent the spins from rotating all together. This procedure, where a random-projection feature is fed to the system every $\Delta t$, is in close analogy with the repeated-encoding prescription in variational quantum circuits, which is known to improve the expressivity of a model~\cite{schuld2021a}.

Shortly after the last pulse of the driving ends, at time $\tau = 30\sigma + M\Delta t $, we get the final encoded state represented by the density matrix $\rhohat(\vb*{x})$. This encoding procedure acts as a non-linear map from the input space of images to the high-dimensional space of $N$-spin mixed quantum states. 

Concerning the non-unitary dynamics due to the openness of the quantum kernel machine, we will consider spin dephasing as the source of decoherence. Within the Lindblad master equation formalism [Eq.~\eqref{eq:mastereq}], this process is described by the jump operators $\hat{A}^j = \hsz^j$, and we consider a uniform dephasing rate for each site $\gamma_i = \gamma,~ \forall i \in \{1,\cdots,N\}$.

Note that while the considered illustrative task involves three classes, it can be reduced to a set of binary classification problems by changing the labels $y \in \{3,6,8\}$ into vector labels of the form $(y_1,y_2,y_3)^T$ with $y_j \in\{-1,1\}^3$.
For example an outcome $(-0.3,-0.2,0.9)$ would correspond to the digit $8$. This ``One-vs-Rest" approach is equivalent to training three binary classifiers, one for each class, and takes the highest output among the three classifiers as a prediction. However, for the sake of simplicity, we will use binary classification notations in the following, and consider that the labels belong to $\{-1,1\}$.

\begin{figure*}[tp!]
    \includegraphics{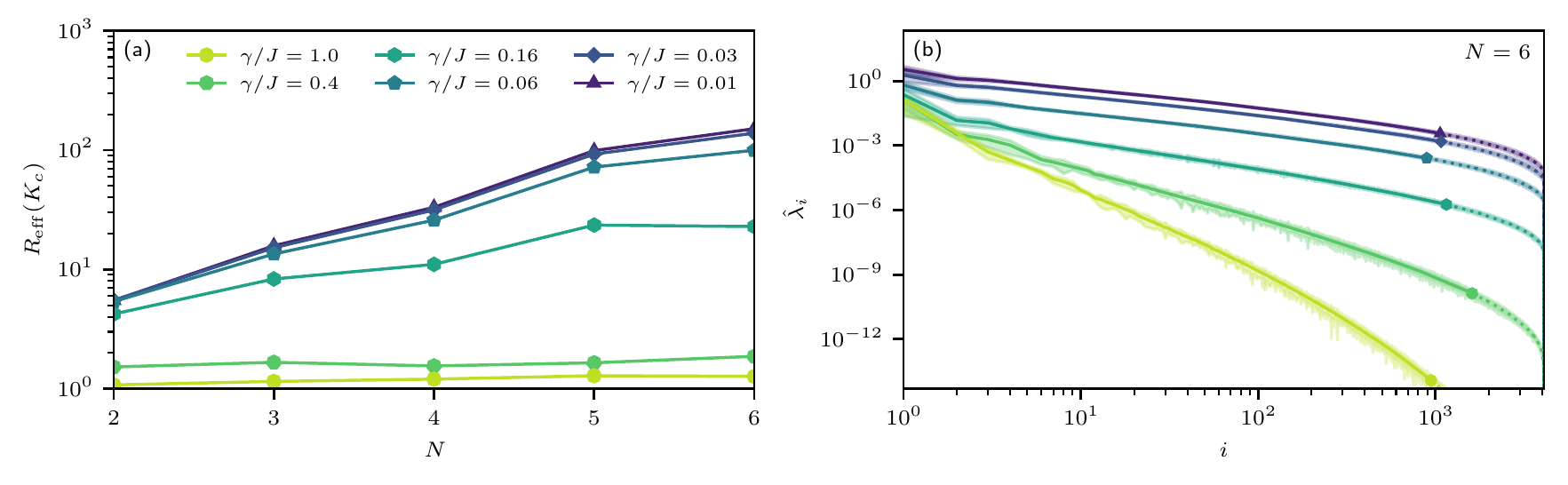}
    \caption{(a) Kernel effective rank for the full tomography decoding as a function of the number of spins $N$ and for different values of the dephasing rate $\gamma$. We have used the empirical representation of the kernel matrix on the training set. The results are averaged over over the same numbers of disorder realizations as for Fig. \ref{fig:full_tomo_results}.
    (b) Kernel empirical spectrum for the full tomography decoding, $N=6$ spins and for different values of the dephasing rate $\gamma$. The markers correspond to the optimal generalization parameter $\lambda$. The curves have been obtained via the kernel empirical representation on the training set. Results are averaged over $25$ disorder configurations. The filled area corresponds to twice the estimated standard error on the averaged value, using the same bootstrap method as for Fig. \ref{fig:full_tomo_results}.
    }
    \label{fig:effective_rank_and_spectra}
    \phantomlabel{a}{fig:effective_rank_and_spectra:a}
    \phantomlabel{b}{fig:effective_rank_and_spectra:b}
\end{figure*}

Regarding the measurements of the system observables, we will consider two measurement protocols:
\begin{enumerate}[(i)]
\item A {\it full tomography} of the output density matrix. In this case we consider that the measurements are made without delay after the end of the encoding, and the extracted features are exactly the components of the generalized Bloch vector $\vb*{\phi}(\vb*{x})$ by considering a complete set of observables.
\item A \textit{time-multiplexing} measurement protocol, where the output is obtained by sequential measurements at different times of a set of local observables.
\end{enumerate}

\subsection{Full tomography}\label{section:complete_tomography_decoding_results}

Any Hermitian operator of the considered spin system can be decomposed on the orthogonal (for the Hilbert-Schmidt inner product) basis of Pauli strings. For a system of $N$ spins, we write this basis
$\{\hat{O}_i ~|~ i=0,\dots,P\}$, with $P=4^N-1$. The corresponding observables are such that
\bea
    &\hat{O}_i = \bigotimes_{k=1}^N\sigmahat^k_{i_k}, \quad i_k \in \{0,1,2,3\};\\
    &\tr\[\hat{O}^{\dagger}_i\hat{O}_j\] = 2^N\delta_{ij}, \quad \forall i,j\, ,
    \label{eq:obs_basis_expansion}
\eea
with $\hat{O}_0 = \hat{\id}$, and thus any observable $\hat{A}$ is decomposed in this basis through the expansion:
\bea
    \hat{A} = \frac{1}{2^N}\(\tr\[\hat{A}\]\hat{\id} + \sum_{i=1}^P\tr\[\hat{O}_i\hat{A}\]\hat{O}_i\).
\eea
The density matrix associated to the input $\vb*{x}$ can also be decomposed into this basis:
\bea
    \rhohat(\vb*{x}) = \frac{1}{2^N}\(\hat{\id}+\sum_{i=1}^{P}\langle \hat{O}_i\rangle_{\vb*{x}} \hat{O}_i\),
    \label{eq:state_basis_expansion}
\eea
and hence any density matrix is uniquely characterized by its associated generalized Bloch vector. For the full-tomography decoding we take these Bloch vectors as the quantum features, which is equivalent to rescaling the quantum kernel function [Eq.~\eqref{eq:quantum_kernel_function}] by a constant factor of $2^N$.

\begin{figure*}[t!]
    \includegraphics{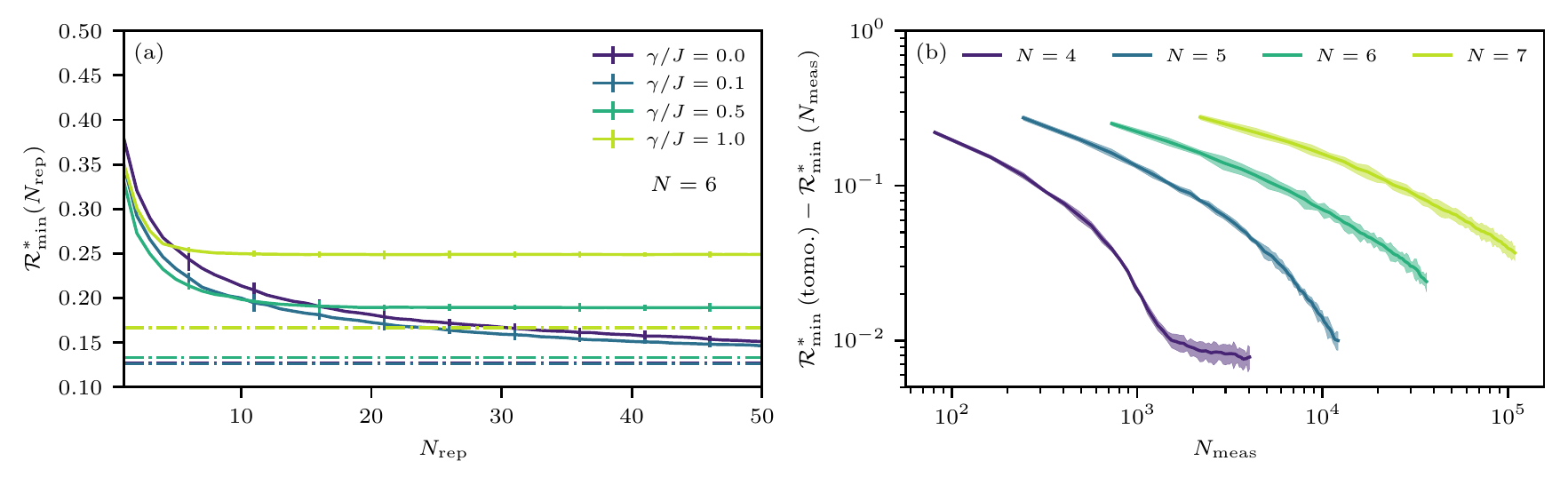}
    \caption{(a) Minimal testing error as a function of the number of repetitions $N_{\rm{rep}}$ of all the local spin measurements for $N=6$ spins and different values of the dephasing noise $\gamma$. The regularization parameter value is chosen to minimize the testing error. The horizontal dashed lines represent the errors obtained using a full tomography decoding for the corresponding dephasing rates. The results are averaged over $50$ disorder realizations. (b) Difference between the testing error of the time-multiplexing decoding and the one for full tomography as a function of the number $N_{\rm{meas}} = 3N N_{\rm{rep}}$ of local measurements performed  for $N_{\rm{rep}} = 50$, dephasing rate $\gamma / J = 0.01$ and different values of the number $N$ of spins. For each realization of the disorder, the regularization parameter value is chosen to minimize the testing error. The results are averaged over $25$ disorder realizations.}
    \label{fig:time_multi_results}
    \phantomlabel{a}{fig:time_multi_results:a}
    \phantomlabel{b}{fig:time_multi_results:b}
\end{figure*}

The encoding method we use leads to embedded states that exhibit entanglement. Fig.~\ref{fig:entanglement_vs_time} shows that the average entanglement negativity quickly increases during the encoding, and then eventually decays at a rate depending on $\gamma$. In parallel, as we see from Fig.~\ref{fig:entropy_vs_time}, there is a finite von Neumann entropy of the system due to mixed character of the state. In Section~\ref{section_numerical_results}, we will show how these processes affect the performances of noisy quantum kernel machines.

\subsection{Time multiplexing measurements}
A simplified and experimentally less demanding decoding is obtained by measuring all the single-site observables (i.e., the three local Pauli spin operators) at different times after the end of the encoding. In the following, we will denote $N_{\rm{rep}}$ the number of repetitions of these measurements. Hence, for a system of $N$ spins, a total number $3N \times N_{\rm{rep}}$ of measurements have been performed after $N_{\rm{rep}}$ repetitions.  We use measurements of the on-site observables for each spin, which correspond to the components of the Bloch vectors of the reduced density matrices on each site. We consider corresponding observables in the Heisenberg picture. The new feature vector $\tilde{\vb*{\phi}}(\vb*{x})$  in the time-multiplexing protocol have entries of the form $\langle B_i(t+k\delta t_m) \rangle_{\vb*{x}}$ with $1\leq i \leq 3N, 1\leq k \leq N_{\rm{rep}}$, where $\delta t_m$ is the time interval between two consecutive measurements.  Similar methods were used in previous works to perform an approximate tomography of the system state \cite{czerwinski2021,dilorenzo2013}. Note that the time-multiplexing procedure can only decrease the model expressive power when compared to the full tomography, as information leaks into the system's environment as the system evolves between successive measurement times (see Appendix \ref{appendix:C_time_multi_expressivity}).

\section{Numerical results}
\label{section_numerical_results}

In this section, we discuss the numerical results on the noisy quantum kernel machines obtained by considering the model spin Hamiltonian, dephasing channels, input encoding via driving, decoding protocol through measurement and the classification task detailed in the previous section.

\subsection{Performances, noise and system size}

The main goal is to determine how the performance of the noisy quantum kernel machine scales with the amount of noise and the number of chain sites, i.e. network nodes. To provide a fair comparison, it is necessary to ensure that the same amount of information is fed into the system for all the system sizes. This is achieved by keeping fixed the number $M$ of projections and resolution of the images. As it will be shown in Section \ref{section:optimized}, the performance can be greatly enhanced when this information bottleneck is lifted and the amount of encoded information is varied. 

The first point to address is the trainability and generalization properties. In Fig.~\ref{fig:full_tomo_results}, we show the dependence of the training and testing errors on the generalization parameter $\lambda$. The curves in (a) are obtained assuming a full tomography and ideal measurements. Panel (b) instead presents the same results, but with imperfect measurements (see caption for more details). In panel (a), the training error (dashed lines) drops to zero as $\lambda \to 0^+$; this is a manifestation of overfitting and indicates that, thanks to the high dimensionality of the quantum feature space, the system is able to completely fit the training data. Instead, the testing error (solid lines and markers) has a minimum value for some optimal value of $\lambda$, which depends on the dephasing rate $\gamma$ (different markers denote different rates). For large enough values of $\lambda$ the testing and training error curves eventually overlap. For increasing $\gamma$ the minimum shifts to vanishing values of $\lambda$. A remarkable result is that the minimal testing error is very little affected by the dephasing rate. 
As shown in Fig.~\ref{fig:full_tomo_results:b}, the situation changes in the presence of imperfect measurements. Indeed, the minimum of the testing error is obtained for a finite value of $\lambda$ even for large values of $\gamma$. Importantly, the minimum error increases with increasing dephasing noise.

In panels (c) and (d) of Fig.~\ref{fig:full_tomo_results}, we report the dependence of the minimal testing error as a function of the number of spins $N$ for increasing values of the dephasing rate. Again, panel (c) corresponds to ideal measurements, while curves in panel (d) are obtained under imperfect measurements. Panel (c) shows that the testing error diminishes as a function of the number of spins and increases with dephasing rate. Note that also for very small dephasing rate the minimal testing error appears to saturate at large system sizes. This is hardly surprising as the input images have been preprocessed and considerably down-sampled. This deliberate choice aims at making the task harder in order to gauge the expressivity of the machine without overloading the input information. As shown in panel (d) of Fig. \ref{fig:full_tomo_results}, by considering imperfect measurements the role of dephasing is dramatically amplified.

As we have described in the analytical discussion in Section~\ref{section:intro_quantum_kernel_and_dissipation}, the quantum kernel spectrum allows us to assess the capacity of our model independently from the specific task one wants to achieve. Fig.~\ref{fig:effective_rank_and_spectra:a} shows the dependence of the quantum kernel's effective rank $R_{\rm{eff}}(\vb{K}_c)$ on the system size and noise strength. For vanishing values of the dephasing rate $\gamma$, we see that this figure of merit first increases exponentially with the number of spins before saturating. For increasing $\gamma$, the effective quantum kernel rank decreases approaching one in the limit of very large $\gamma$.

The same behavior is observed in the empirical spectrum in Fig.~\ref{fig:effective_rank_and_spectra:b} as the noise rate is varied. For increasing values of $\gamma$, we observe a faster decrease of the empirical kernel eigenvalues as a function of the eigenvalue number. For comparison, we have indicated with markers the largest eigenvalue below the optimal generalization parameter.  This gives a rough estimate of the number of kernel eigenfunctions required to correctly approximate the target function. Note that in the context of imperfect measurements the generalization parameter is bounded from below, and hence some of the kernel eigenfunctions becomes out-of-reach. This shows a clear link between the kernel eigenvalues and the expressivity of the machine.

The results discussed relied on full tomography. As we have explained in Section~\ref{section:driven_dissipative_spins_model}, it is possible to design a simplified and less expensive measurement protocol based on a time-multiplexing procedure where a set of local spin observables are measured at $N_{\rm{rep}}$ different times. The results obtained with such an approach are summarized in Fig.~\ref{fig:time_multi_results}. As appears from Fig.~\ref{fig:time_multi_results:a}, by increasing the number of repetitions $N_{\rm{rep}}$ the error diminishes. For small enough values of dephasing $\gamma$, the error 
converges to the value in the ideal case of full-tomography. For increasing $\gamma$, however, the saturating value departs from the ideal one given by full tomography, showing that the time-multiplexing expressivity deteriorates more than that of the full tomography for larger noise.
This trend is further elucidated in Fig.~\ref{fig:time_multi_results:b}, where the difference between the time-multiplexing error and the full-tomography error is reported as a function of the total number of measured observables. By increasing the number of spins  and hence the dimension of the Hilbert space for a given dephasing rate, the required number of repetitions increases.


\label{section:time_multi_decoding_results}

\subsection{Optimizing the encoding}
\label{section:optimized}

\begin{figure}[t!]
    \centering
    \includegraphics[width=\linewidth]{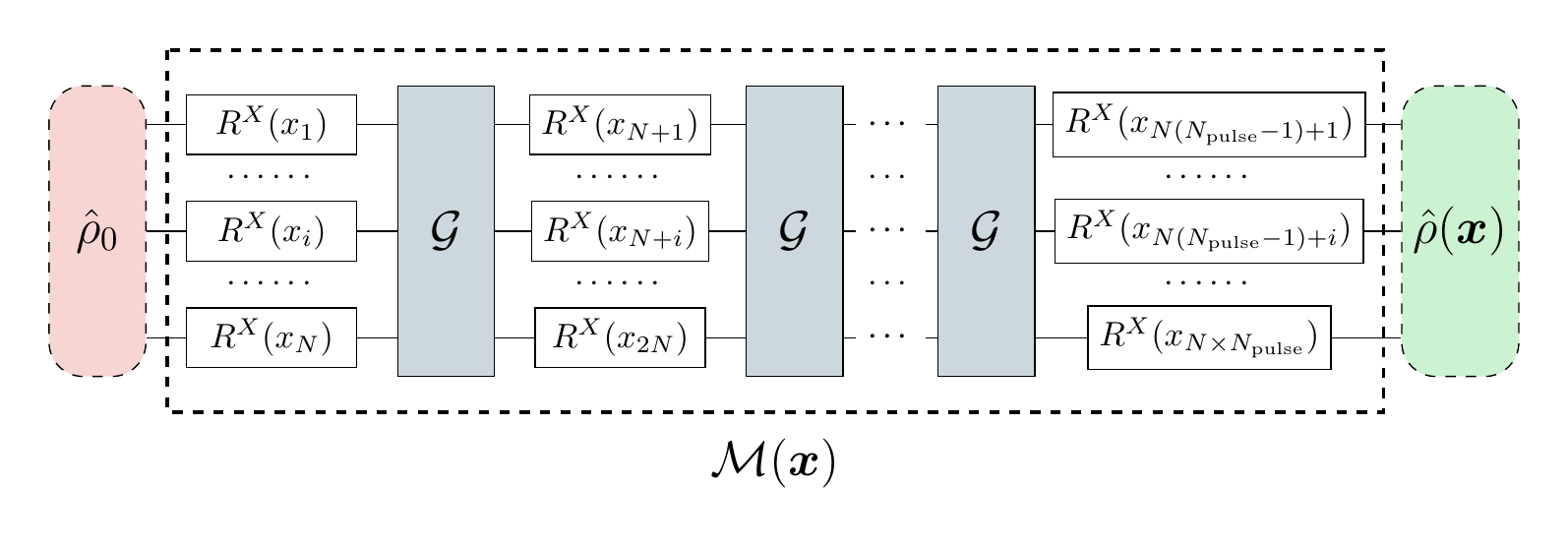}
    \caption{Equivalent circuit of the encoding with the information bottleneck removed. Instead of injecting a single random projection feature $x_i$ per time step as represented in Fig.~\ref{fig:circuit}, where a total number of $M = \npulse$ random projection features are fed into the kernel machine, here we inject $N$ random projection features in each time step, with a total number of $M=N\times\npulse$ random projection features injected by the end of the encoding sequence.}
    \label{fig:circuit_new}
\end{figure}

\begin{figure}[t!]
    \centering
    \includegraphics[width=\linewidth]{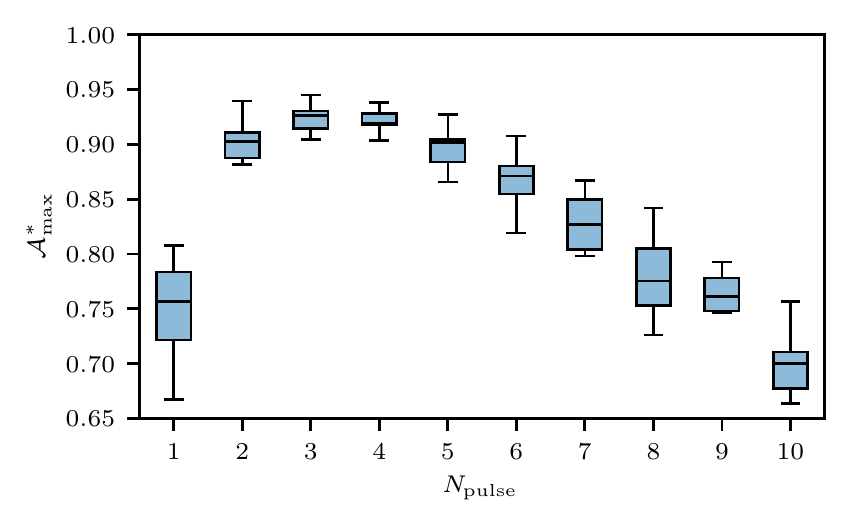}
    \caption{Optimal testing accuracy as a function of the number of driving pulses, for the encoding presented on Fig.~\ref{fig:circuit_new}, $N=6$ spins and $\gamma/J=0.01$. The regularization parameter is chosen as to maximize the testing accuracy. Note that for this encoding the number of features $M$ yielded by the preprocessing is $M = N\times \npulse$. The results are shown for $10$ realizations of the disorder on both the preprocessing and the system parameters. The maximal accuracy obtained is $94.5\%$ for $\npulse = 3$. The boxes extend from the first to the third quartile of the distributions, the middle line indicates the median and the wiskers indicates the extreme values of the distributions.}
    \label{fig:fid_vs_M_new_encoding}
\end{figure}

In this section we investigate an alternative encoding scheme for which the amount of information fed to the system scales with the system size. The embedding studied in the previous sections involved a set of $M = \npulse$ random-projection features derived from the down-sampled images. Here we derive a number $M = N\times \npulse$ of such features and split them in $N$ sequences of $\npulse$ features, which we use to drive the $N$ sites. In particular, the driving sequences sent to different sites are unique, while for the previous encoding those sequences were proportional to each other. The new encoding procedure is presented in Fig.~\ref{fig:circuit_new} in the form of its equivalent circuit. In Fig.~\ref{fig:fid_vs_M_new_encoding} we report the evolution of the performances given by this new encoding as a function of the number of driving pulses. As the number of pulses rises the corresponding number of encoded features $M=N\times \npulse$ increases and so does the amount of encoded information. The corresponding maximal testing accuracy reaches an optimum of $94.5\%$ for $\npulse = 3$. For large number of pulses $\npulse$ the performances drop. This effect is due to the fact that for such parameters the transformations yielded by the encoding are poorly adapted to the task at hand, i.e. the kernel and the target function become less ``aligned". Note that the performance is very sensitive to both the encoding method and the physical system parameters. While controlling the physical parameters might be hard, it appears that a careful design of the encoding procedure can significantly boost the performances. This makes the research of tailored encoding procedures a promising avenue of research.

\section{Conclusion}
\label{section:conclusion}
In this work, we have presented a quantum machine-learning model based on the quantum-kernel paradigm. Within the formalism of kernel theory, we have characterized the expressivity and generalization capacity of this model. We have linked the relevant figure of merits to the spectrum of the associated centered quantum kernel. In particular, we presented an upper bound on the generalization error involving the average purity of quantum states representing the data to classify. This upper-bound shows that dissipation and decoherence act as a regularization for the quantum kernel machines. By considering an illustrating example of a driven-dissipative spin chain as the noisy quantum kernel machine, we have shown how the expressivity and generalization capacity are controlled by both the dephasing rate and by experimental uncertainties on the measurements. Moreover, we have shown how the performances of the noisy quantum kernel machines are modified when the full-tomography measurement protocol is replaced by a time-multiplexing procedure requiring only local observables, and how the openness of the system mitigates the efficiency of this protocol.
We observed a qualitative improvement in the processing performance of our model when going from a scenario where the system is fed a constant amount of information to one where the inputs are encoded at a finite information rate that scales extensively with the system size. How to design tailored encoding strategies able to harness the full power of quantum kernel machines remains an open question. In particular, investigating encoding schemes that would allow to inject information at a rate scaling exponentially in the system size seems promising.
The concepts presented here and the unavoidable role of the decoherence in any realistic physical system are relevant for a wide range of quantum machine-learning models, ranging from quantum extreme-learning machines to quantum neural networks.

\begin{acknowledgments}
     We would like to thank Hugo Tschirhart for help at the early stages of this project. This work was supported by the FET FLAGSHIP Project PhoQuS (grant agreement ID: 820392), by Region Île-de-France in the framework of DIM SIRTEQ, and the French agency ANR through the grants NOMOS (ANR-18-CE24-0026) and TRIANGLE (ANR-20-CE47-0011).
    We also acknowledge access to the high performance computation center TGCC of the French national computational facility GENCI under the project 2021-A0100512462.
    
\end{acknowledgments}

\appendix
\section{Expressivity and generalization for noisy quantum kernel}
\label{appendix:A_expressivity_and_generalization_dissipative_quantum_kernel}

\subsection{Expressivity and kernel effective rank}
\label{appendix:A1_kernel_expressivity_and_effective_rank}

To measure the ability of a kernel $k$ to learn a function $y(\vb*{x})$, we have introduced in Eq. \eqref{eq:alignment} the {\it kernel target alignment} $A(k,y)$. We then defined the {\it kernel effective rank} $R_\mathrm{eff}$ by considering a set of orthonormal basis functions $\{g_i\}$, that gives the following equalities:
\bea
    \sqrt{R_{\rm{eff}}(k)}=&\sum_{j}A(k,g_j)\\
    =& \dfrac{1}{(\sum_i \lambda_i^2)^{1/2}}\sum_j \sum_i \lambda_i \expectt{p}{\psi_i(\vb*{x})g_j(\vb*{x})}^2\\
    =& \dfrac{1}{(\sum_i \lambda_i^2)^{1/2}}\sum_i \lambda_i \expectt{p}{\psi_i(\vb*{x})^2}\\
    =& \dfrac{\sum_i \lambda_i}{(\sum_i \lambda_i^2)^{1/2}}\, .
\eea
Note that the final expression concerns only the spectrum of the kernel and is independent of the choice of the basis functions $\{g_i\}$.
 From the Cauchy-Schwarz inequality, we have 
\bea
    R_{\rm{eff}}(k)\leq \vert{}\{ \lambda_i\neq 0 \}\vert{} \, ,
\eea
where the equality is attained if and only if all non-zero eigenvalues of the kernel are equal. Therefore, it provides information about the flatness of the spectrum of the kernel. 

Given a training sample of size $\ntrain$, the kernel spectrum can be empirically computed using the $\ntrain\times \ntrain$ kernel matrix $\vb{K}$ associated to the kernel $k$, whose entries are $K_{ij} = k(\vb*{x}_i,\vb*{x}_j)$. The eigenvalues $\lambda_i$ of the kernel $k$ can then be approximated by those of the matrix $\vb{K}/\ntrain$ ~\cite{Williams01usingthe}.
For the centered quantum kernel $k_c$ with the associated kernel matrix $\vb{K}_c$, we can compute the effective rank empirically as:
\bea\label{eq:reff_app}
    \sqrt{R_{\rm{eff}}(\vb{K}_c)} = \frac{\tr\[\vb{K}_c\]}{\sqrt{\tr\[\vb{K}_c^2\]}}
    = \frac{\sum_i\hat{\lambda}_i}{\sqrt{\sum_i\hat{\lambda}_i^2}}\, ,
\eea
where the $\hat{\lambda}_i$ are the empirical eigenvalues
\footnote{The hat symbol is used for estimators such as the empirical eigenvalues, as customary in statistical theory. The hat must not confused with the one used for the quantum operators}.

The numerator can be expressed using the empirical kernel eigenobservables. In order to keep light notations we use the same notations as in the main text, i.e. the empirical kernel eigenobservables are denoted $\hat{E}_i$. Whether this notation refers to the exact or the empirical observable should be clear from the context. We have for the numerator:
\bea\label{eq:sum_lambda}
    \sum_i\hat{\lambda}_i &= \expectt{\hat{p}}{\sum_i\delta\langle\hat{E}_i\rangle_{\vb*{x}}^2}\\
    &= \expectt{\hat{p}}{\sum_i\tr\[\delta\rhohat(\vb*{x})\hat{E}_i\]^2}\, .
\eea
Since $\tr[\delta\rhohat(\vb*{x})] = 0$, $\delta\rhohat(\vb*{x})$ can be decomposed onto the eigenobservable basis $\{\hat{E}_i\}$ through the expression:
\bea
\delta\rhohat(\vb*{x}) = \sum_i \tr\[\delta\rhohat(\vb*{x})\hat{E}_i\]\hat{E}_i \, .
\eea
Consequently, the squared Hilbert-Schmidt norm reads:
\bea
\tr\[\delta\rhohat(\vb*{x})^2\] = \sum_i \tr\[\delta\rhohat(\vb*{x})\hat{E}_i\]^2\, .
\eea
Eq. \eqref{eq:sum_lambda} therefore becomes:
\bea\label{eq:kernel_trace_app}
\sum_i\hat{\lambda}_i &= \expectt{\hat{p}}{\tr\[\delta\rhohat(\vb*{x})^2\]}\\
&= \tr\[\expectt{\hat{p}}{\(\rhohat(\vb*{x}) - \expectt{\hat{p}}{\rhohat(\vb*{x})}\)^2}\]\\
&= \expectt{\hat{p}}{\tr\[\rhohat(\vb*{x})^2\]} - \tr\[\expectt{\hat{p}}{\rhohat(\vb*{x})}^2\] \, ,
\eea
giving Eq.~\eqref{eq:kernel_trace_and_purity} in the main text (as the same relation holds between the true eigenvalues $\lambda_i$ and the distribution $p$). This quantity can also be written in terms of the measured observables (note that $\mathcal{O} = \mathcal{B}$ for a quantum kernel):
\bea
    \sum_i\hat{\lambda}_i =& \dfrac{\tr\[\vb{K}_c\]}{\ntrain}\\
    =& \dfrac{1}{\ntrain}\sum_i k_c(\vb*{x}_i,\vb*{x}_i)\\
    =& \dfrac{1}{\ntrain}\sum_i\sum_k\delta\phi_k(\vb*{x}_i)\delta\phi_k(\vb*{x}_i)\\
    =& \sum_k\left( \dfrac{1}{\ntrain}\sum_i\delta\phi_k(\vb*{x}_i)\delta\phi_k(\vb*{x}_i)\right)\\
    =& \sum_k \expectt{\hat{p}}{\delta\phi_k(\vb*{x})^2}\\
    =& \sum_k \expectt{\hat{p}}{\delta\meanq{\hat{O}_k}_{\vb*{x}}^2}\\
    =& \sum_k \mathrm{Var}_{\hat{p}}\left[\meanq{\hat{O}_k}_{\vb*{x}}  \right]\, .
\eea

Similarly, in the denominator of Eq.~\eqref{eq:reff_app}, we get:
\bea
\sum_i \hat{\lambda}_i^2 &= \frac{\tr\[\vb{K}^2_c\]}{\ntrain^2}\\
&= \frac{1}{\ntrain^2}\sum_{i,j} k_c\(\vb*{x}_i, \vb*{x}_j\)^2 \\
&= \frac{1}{\ntrain^2}\sum_{i,j} \(\sum_k \delta\phi_k(\vb*{x}_i)\delta\phi_k(\vb*{x}_j)\)^2 \\
&= \sum_{k,l}\(\frac{1}{\ntrain}\sum_i \delta\phi_k(\vb*{x}_i)\delta\phi_l(\vb*{x}_i)\)^2\\
&= \sum_{k,l}\expectt{\hat{p}}{\delta\meanq{\hat{O}_k}_{\vb*{x}}\delta\meanq{\hat{O}_l}_{\vb*{x}}}^2 \\
&= \sum_{k,l}\mathrm{Cov}_{\hat{p}}\[\meanq{\hat{O}_k}_{\vb*{x}},\meanq{\hat{O}_l}_{\vb*{x}}\]^2\, .
\eea

Finally, we get the general expression: 
\bea
     \sqrt{R_{\rm{eff}}(\vb{K}_c)}=\frac{\sum_{i=1}^P \mathrm{Var}_{\hat{p}}\[\meanq{\hat{O}_i}_{\vb*{x}}\]}{\(\sum_{i,j=1}^P\mathrm{Cov}_{\hat{p}}\[\meanq{\hat{O}_i}_{\vb*{x}},\meanq{\hat{O}_j}_{\vb*{x}}\]^2\)^{\frac{1}{2}}}\, .
\eea
Note that this relation also holds for the true (non-empirical) effective rank $R_\mathrm{eff}(k_c)$ provided that the variances and the covariances are taken with respect to the true probability distribution $p$ instead of the empirical one $\hat{p}$.

\subsection{Generalization and Rademacher complexity}
\label{appendix:A2_generalization_bound_kernel}

Here we give the detailed derivation of Eq. \eqref{eq:generalization_error_and_dissipation} using methods of statistical learning theory applied to the specific case of a noisy centered quantum kernel \cite{mohri2018foundations}. 

In the standard setup of statistical learning, the inputs $\vb*{x}\in\mathcal{X}$ are considered as a random variable following a probability distribution $p(\vb*{x})$. We define the target function $y:\mathcal{X}\mapsto\mathcal{Y}$ that assigns to each input its right label.  We will consider the case of a binary classification, for which $\mathcal{Y} = \{-1,1\}$. In practice the true distribution $p$ of the inputs is unknown and during the training we only have access to a finite training dataset $\mathcal{S} = \{(\vb*{x}_i, y_i)\mid i=1,\dots,\ntrain\}$. The elements of the dataset are considered as realizations of a set of independent and identically distributed random variables following $p$. The empirical distribution associated to this training set is given by:
\bea
\hat{p}(\vb*{x}) = \frac{1}{\ntrain}\sum_{i=1}^{\ntrain} \delta(\vb*{x}-\vb*{x}_i) \,.
\eea
We rely on this empirical distribution to evaluate expectations of any function $f(\vb*{x})$, namely:
\bea
\expectt{p}{f(\vb*{x})} = \int_{\mathcal{X}}f(\vb*{x})p(\vb*{x})d\vb*{x} \,.
\eea
The expectation value is approximated by its empirical counterpart:
\be
\expectt{\hat{p}}{f(\vb*{x})} = \int_{\mathcal{X}}f(\vb*{x})\hat{p}(\vb*{x})d\vb*{x} = \frac{1}{\ntrain}\sum_{i=1}^{\ntrain} f(\vb*{x}_i) \, 
\ee
 A common question in statistical learning is to know how a model trained on a given set of data will perform on any other set of unseen data. For a binary classification task with balanced data one can use the accuracy as measure of the model performance. Given a trial function $f(\vb*{x}) = \vb*{w}^T\vb*{\phi}(\vb*{x})$ that has been optimized using the training set $\mathcal{S}$, we define the corresponding prediction function as $\tilde{f}(\vb*{x}) = \mathrm{sign}[f(\vb*{x})]$. An input $\vb*{x}$ is correctly classified if $\tilde{f}(\vb*{x}) = y(\vb*{x})$. The true accuracy $\mathcal{A}^{*}(f)$ is defined as the probability that any input in $\mathcal{X}$ is correctly classified by $\tilde{f}$:
\bea
\mathcal{A}^{*}(f) &= \expectt{p}{\id_{y(\vb*{x})=\tilde{f}(\vb*{x})}} = \expectt{p}{\id_{y(\vb*{x})f(\vb*{x})\geq0}}\\
&= 1 - \expectt{p}{\id_{y(\vb*{x})f(\vb*{x})\leq0}} = 1 - \mathcal{R}^*(f) \, ,
\eea
where we define the risk (also called error or inaccuracy) as $\mathcal{R}^*(f) = 1-\mathcal{A}^*(f)$. The corresponding empirical quantities $\mathcal{A}(f)$ and $\mathcal{R}(f)$ are defined in an analogous way using the empirical distribution $\hat{p}$ instead of $p$. The ability to perform well on new data is measured by the generalization error:
\bea
\mathcal{E}(f) = \mathcal{R}^*(f) - \mathcal{R}(f) \,.
\eea
Statistical learning theory provides probabilistic upper-bounds on the generalization error depending on the type of task at hand and on the specific model used to tackle it. In order to find such an upper bound for a binary classification tasks, it is convenient to consider a relaxed version of the risk, the $\eta$-margin-risk $\mathcal{R}_\eta(f)$ defined in the main text. 
The upper bound on the generalization properties involves the empirical Rademacher complexity of a class of trial functions $\mathcal{H}$ with respect to the training sample $\mathcal{S}$. It is defined as:
\bea
    \mathfrak{R}_{\mathcal{S}}\(\mathcal{H}\) = \mathbb{E}_{\vb*{\sigma}}\[ \sup_{f\in\mathcal{H}}\frac{1}{\ntrain}\sum_{i=1}^{\ntrain}\sigma_i f(\vb*{x}_i)\] \,
\eea
where $\vb*{\sigma}$ is a vector of Rademacher variables that are discrete, independent and identically distributed following a uniform law over $\{-1,1\}$. The Rademacher complexity measures the ability of a hypothesis class $\mathcal{H}$ to fit noise, and as such it is a measure of the expressivity of $\mathcal{H}$. We now give a  upper-bound on the generalization error (Theorem 5.8 in \cite{mohri2018foundations}):
\begin{theorem}[]
Let $\mathcal{H}$ be a set of trial functions and $\eta>0$. Then $\forall \delta>0$, with probability at least $1-\delta$, we have $\forall f \in \mathcal{H}$:
$$\mathcal{R}^*(f) \leq \mathcal{R}_{\eta}(f) + \frac{2}{\eta}\mathfrak{R}_{\mathcal{S}}(\mathcal{H}) + 3 \sqrt{\frac{\mathrm{log}(\frac{2}{\delta})}{2\ntrain}}$$
\end{theorem}
This upper-bound can be specialized to the case of kernel methods where the hypothesis class is the RKHS of a kernel $k$. In this the Rademacher complexity is upper bounded by a quantity that depends only on the trace of the empirical kernel matrix $\vb{K}$ (Theorem 6.12 in \cite{mohri2018foundations}):
\begin{theorem}[]
Let $\mathcal{H}$ by the RKHS associated to a given kernel $k$. For $\Lambda\geq 0$ consider the set of hypothesis functions $\mathcal{H}_{\Lambda} = \{f: x \mapsto \vb*{w}^T\delta\vb*{\phi}(x), \quad \norm{\vb*{w}}^2\Lambda\leq 1\}\subseteq\mathcal{H}$. Then we have: $$\mathfrak{R}_{\mathcal{S}}\(\mathcal{H}_{\Lambda}\) \leq \frac{1}{\ntrain}\sqrt{\frac{\tr\[\vb{K}\]}{\Lambda}}\, .$$
\end{theorem}
Injecting this result in the previous upper bound, we get the desired result. In particular, using the centered noisy quantum kernel $k_c$ and Eq.~\eqref{eq:kernel_trace_app}, we get Eq. \eqref{eq:generalization_error_and_dissipation}.

\section{Intercept and kernel centering}
\label{appendix:B_intercept_and_kernel_centering}

The initial optimization problem is to find a weight vector $\vb*{w} = (b,w_1,\dots,w_P)^T$ that minimizes the regularized loss function of Eq. \eqref{eq:L2loss_with_reg}. 
We slightly change our notation and drop the first constant term of the embedding map $\vb*{\phi}(\vb*{x})$ to explicitly seperate the bias $b$ from the weight $\vb*{w}$, so that the loss can be rewritten:
\bea
\mathcal{L}\left(\vb*{w},b \mid \mathcal{S}\right) =&\frac{1}{2\ntrain}\sum_{i=1}^{\ntrain}\(y_i - \vb*{w}^T\vb*{\phi}(\vb*{x}_i)-b\)^2\\&+\frac{\lambda}{2}\norm{\vb*{w}}_2^2\, .\\
\eea
The optimal intercept $b$ is found by imposing $\frac{\partial \mathcal{L}}{\partial b} = 0$. The solution reads:
\bea
b^* = \frac{1}{\ntrain}\sum_{i=1}^{\ntrain} y_i - \vb*{w}^T\(\frac{1}{\ntrain}\sum_{i=1}^{\ntrain}\vb*{\phi}(\vb*{x}_i)\) \,.
\eea
We see that the optimal intercept consists of two terms: one that has the effect of centering the labels, while the other centers the features. Assuming the dataset we use are balanced, we have $\sum_{i=1}^{\ntrain} y_i \simeq 0$ Plugging back the optimal intercept into the previous regularized loss function,  we get a new effective loss:
\bea
\mathcal{L}^{\ast}(\vb*{w} \mid \mathcal{S}) =&\frac{1}{2\ntrain}\sum_{i=1}^{\ntrain}\(y_i - \vb*{w}^T\(\vb*{\phi}(x_i)-\expectt{\hat{p}}{\vb*{\phi}}\)\)^2\\
&+\frac{\lambda}{2}\norm{\vb*{w}}_2^2 \, .
\eea
If the data are not balanced one can simply replace the labels $y_i$ by their centered counterpart $y_i' = y_i - \frac{1}{\ntrain}\sum_{i=1}^{\ntrain} y_i$ such that $\sum_{i=1}^{\ntrain} y_i' = 0$.
Note that this might lead to issues when using the accuracy metric with unbalanced labels. This issue can be fixed, e.g. by changing the metric used for a balanced one.
Thus, working with the quantum kernel without regularizing the intercept term is equivalent to working with the centered kernel and centered labels.

\section{Time-multiplexing and model expressivity}
\label{appendix:C_time_multi_expressivity}

The maximal class of trial functions $\mathcal{H}_{\rm{full}}$ (see section \ref{section:intro_quantum_kernel_and_dissipation}) associated to a given embedding is obtained by performing a complete tomography of the embedded quantum states $\rhohat(\vb*{x})$ right after the end of the encoding procedure. The system evolution after time $\tau$ according to a Lindblad master equation with a constant Hamiltonian and disspator for a given duration $\delta t_m$ can be expressed into a set of Kraus operators $\{\hat{W}_i\}$ \cite{breuer2007} satisfying:
\bea
    \sum_i \hat{W}_i^{\dagger}\hat{W}_i = \hat{\id} \,.
\eea
The evolved density matrices $\rhohat(\vb*{x}; \delta t_m)$ are given by:
\bea
\rhohat(\vb*{x}; \delta t_m) = \sum_i \hat{W}_i \rhohat(\vb*{x}) \hat{W}_i^{\dagger} \,.
\eea
In the Heisenberg picture, the observables evolve in time following an adjoint master equation \cite{breuer2007}. Hence,  we can see the non-unitary evolution of the open quantum system as a simple change in the set of observables that are measured on the state $\rhohat(\vb*{x})$. Suppose that we want to measure observables from the orthonormal basis introduced in section \ref{section:decoding_methods} after the previous evolution. We define the $(P+1)\times(P+1)$ matrix $\vb{\Xi}$ whose elements are:
\bea
\Xi_{kl} = \tr\[\sum_i\hat{W}_i^{\dagger}\hat{O}_k\hat{W}_i\hat{O}_l\] \,.
\eea
The measurement at time $\tau + \delta t_m$ of the observable $\hat{O}_l$ can now be expressed using the decomposition in Eq. \eqref{eq:state_basis_expansion} and the elements of $\vb{\Xi}$ as:
\bea
\tr\[\rhohat(\vb*{x};\delta t_m)\hat{O}_l\] = \frac{1}{2^N}\(\Xi_{0l}+\sum_{k}\tr\[\rhohat(\vb*{x})\hat{O}_k\]\Xi_{kl}\) \,.
\eea
Thus the embedding map $\vb*{\phi}(\vb*{x})$ is transformed by the non-unitary evolution during $\delta t_m$ and becomes:
\bea
\vb*{\phi}(\vb*{x};\delta t_m) = \vb{\Xi}\vb*{\phi}(\vb*{x}) \,.
\eea
Assuming we only make measurements on a subset of the basis $\{\hat{O}_i\}$, then we can write for the feature vector:
\bea
\vb*{\phi}(\vb*{x};\delta t_m) = \vb{D}\vb{\Xi}\vb*{\phi}(\vb*{x}) \,,
\eea
where $\vb*{D}$ is a diagonal $(P+1)\times(P+1)$ matrix whose diagonal entries $i$ are $1$ if $\hat{O}_i$ is measured and $0$ otherwise. When we repeat the measurements at different times, we can stack the previous vectors at each time steps. For $N_{\rm{rep}}$ repetitions, we denote $\vb{\Lambda}$ the $N_{\rm{rep}}(P+1)\times(P+1)$ matrix of the form:
\bea
\vb{\Lambda} &= \begin{pmatrix}
\vb{D}\vb{\Xi}\\
\vb{D}\vb{\Xi}^2\\
\vdots \\
\vb{D}\vb{\Xi}^{N_{\rm{rep}}} \\
\end{pmatrix}\, .
\eea
The final vector reads:
\bea
\vb*{\tilde{\phi}}(\vb*{x}) = \vb{\Lambda}\vb*{\phi}(\vb*{x}) \, .
\eea
Hence, by performing repeated measurements in-between non-unitary evolutions amount to performing a restricted number of measurements on the encoded states $\rhohat(\vb*{x})$ at time $\tau$.  This implies that the time-multiplexing decoding lowers the model expressivity. The difference between the models obtained from the full tomography and the time-multiplexing decoding is encapsulated in the matrix $\vb*{\Lambda}$.

\bibliography{bib.bib}

\begin{thebibliography}{56}%
\makeatletter
\providecommand \@ifxundefined [1]{%
 \@ifx{#1\undefined}
}%
\providecommand \@ifnum [1]{%
 \ifnum #1\expandafter \@firstoftwo
 \else \expandafter \@secondoftwo
 \fi
}%
\providecommand \@ifx [1]{%
 \ifx #1\expandafter \@firstoftwo
 \else \expandafter \@secondoftwo
 \fi
}%
\providecommand \natexlab [1]{#1}%
\providecommand \enquote  [1]{``#1''}%
\providecommand \bibnamefont  [1]{#1}%
\providecommand \bibfnamefont [1]{#1}%
\providecommand \citenamefont [1]{#1}%
\providecommand \href@noop [0]{\@secondoftwo}%
\providecommand \href [0]{\begingroup \@sanitize@url \@href}%
\providecommand \@href[1]{\@@startlink{#1}\@@href}%
\providecommand \@@href[1]{\endgroup#1\@@endlink}%
\providecommand \@sanitize@url [0]{\catcode `\\12\catcode `\$12\catcode
  `\&12\catcode `\#12\catcode `\^12\catcode `\_12\catcode `\%12\relax}%
\providecommand \@@startlink[1]{}%
\providecommand \@@endlink[0]{}%
\providecommand \url  [0]{\begingroup\@sanitize@url \@url }%
\providecommand \@url [1]{\endgroup\@href {#1}{\urlprefix }}%
\providecommand \urlprefix  [0]{URL }%
\providecommand \Eprint [0]{\href }%
\providecommand \doibase [0]{http://dx.doi.org/}%
\providecommand \selectlanguage [0]{\@gobble}%
\providecommand \bibinfo  [0]{\@secondoftwo}%
\providecommand \bibfield  [0]{\@secondoftwo}%
\providecommand \translation [1]{[#1]}%
\providecommand \BibitemOpen [0]{}%
\providecommand \bibitemStop [0]{}%
\providecommand \bibitemNoStop [0]{.\EOS\space}%
\providecommand \EOS [0]{\spacefactor3000\relax}%
\providecommand \BibitemShut  [1]{\csname bibitem#1\endcsname}%
\let\auto@bib@innerbib\@empty
\bibitem [{\citenamefont {LeCun}\ \emph {et~al.}(2015)\citenamefont {LeCun},
  \citenamefont {Bengio},\ and\ \citenamefont {Hinton}}]{lecun2015}%
  \BibitemOpen
  \bibfield  {author} {\bibinfo {author} {\bibfnamefont {Yann}\ \bibnamefont
  {LeCun}}, \bibinfo {author} {\bibfnamefont {Yoshua}\ \bibnamefont {Bengio}},
  \ and\ \bibinfo {author} {\bibfnamefont {Geoffrey}\ \bibnamefont {Hinton}},\
  }\bibfield  {title} {\enquote {\bibinfo {title} {Deep learning},}\ }\href
  {\doibase 10.1038/nature14539} {\bibfield  {journal} {\bibinfo  {journal}
  {Nature}\ }\textbf {\bibinfo {volume} {521}},\ \bibinfo {pages} {436--444}
  (\bibinfo {year} {2015})}\BibitemShut {NoStop}%
\bibitem [{\citenamefont {Goodfellow}\ \emph {et~al.}(2016)\citenamefont
  {Goodfellow}, \citenamefont {Bengio},\ and\ \citenamefont
  {Courville}}]{goodfellow2016}%
  \BibitemOpen
  \bibfield  {author} {\bibinfo {author} {\bibfnamefont {Ian}\ \bibnamefont
  {Goodfellow}}, \bibinfo {author} {\bibfnamefont {Yoshua}\ \bibnamefont
  {Bengio}}, \ and\ \bibinfo {author} {\bibfnamefont {Aaron}\ \bibnamefont
  {Courville}},\ }\href@noop {} {\emph {\bibinfo {title} {Deep
  {{Learning}}}}},\ edited by\ \bibinfo {editor} {\bibfnamefont {Francis}\
  \bibnamefont {Bach}},\ Adaptive {{Computation}} and {{Machine Learning}}
  Series\ (\bibinfo  {publisher} {{MIT Press}},\ \bibinfo {address}
  {{Cambridge, MA, USA}},\ \bibinfo {year} {2016})\BibitemShut {NoStop}%
\bibitem [{\citenamefont {Strubell}\ \emph {et~al.}(2019)\citenamefont
  {Strubell}, \citenamefont {Ganesh},\ and\ \citenamefont
  {McCallum}}]{strubell2019}%
  \BibitemOpen
  \bibfield  {author} {\bibinfo {author} {\bibfnamefont {Emma}\ \bibnamefont
  {Strubell}}, \bibinfo {author} {\bibfnamefont {Ananya}\ \bibnamefont
  {Ganesh}}, \ and\ \bibinfo {author} {\bibfnamefont {Andrew}\ \bibnamefont
  {McCallum}},\ }\bibfield  {title} {\enquote {\bibinfo {title} {Energy and
  policy considerations for deep learning in {NLP}},}\ }in\ \href {\doibase
  10.18653/v1/P19-1355} {\emph {\bibinfo {booktitle} {Proceedings of the 57th
  Annual Meeting of the Association for Computational Linguistics}}}\ (\bibinfo
   {publisher} {Association for Computational Linguistics},\ \bibinfo {year}
  {2019})\ pp.\ \bibinfo {pages} {3645--3650}\BibitemShut {NoStop}%
\bibitem [{\citenamefont {Tanaka}\ \emph {et~al.}(2019)\citenamefont {Tanaka},
  \citenamefont {Yamane}, \citenamefont {H{\'e}roux}, \citenamefont {Nakane},
  \citenamefont {Kanazawa}, \citenamefont {Takeda}, \citenamefont {Numata},
  \citenamefont {Nakano},\ and\ \citenamefont {Hirose}}]{tanaka2019}%
  \BibitemOpen
  \bibfield  {author} {\bibinfo {author} {\bibfnamefont {Gouhei}\ \bibnamefont
  {Tanaka}}, \bibinfo {author} {\bibfnamefont {Toshiyuki}\ \bibnamefont
  {Yamane}}, \bibinfo {author} {\bibfnamefont {Jean~Benoit}\ \bibnamefont
  {H{\'e}roux}}, \bibinfo {author} {\bibfnamefont {Ryosho}\ \bibnamefont
  {Nakane}}, \bibinfo {author} {\bibfnamefont {Naoki}\ \bibnamefont
  {Kanazawa}}, \bibinfo {author} {\bibfnamefont {Seiji}\ \bibnamefont
  {Takeda}}, \bibinfo {author} {\bibfnamefont {Hidetoshi}\ \bibnamefont
  {Numata}}, \bibinfo {author} {\bibfnamefont {Daiju}\ \bibnamefont {Nakano}},
  \ and\ \bibinfo {author} {\bibfnamefont {Akira}\ \bibnamefont {Hirose}},\
  }\bibfield  {title} {\enquote {\bibinfo {title} {Recent advances in physical
  reservoir computing: {{A}} review},}\ }\href {\doibase
  10.1016/j.neunet.2019.03.005} {\bibfield  {journal} {\bibinfo  {journal}
  {Neural Networks}\ }\textbf {\bibinfo {volume} {115}},\ \bibinfo {pages}
  {100--123} (\bibinfo {year} {2019})}\BibitemShut {NoStop}%
\bibitem [{\citenamefont {Huang}\ \emph {et~al.}(2006)\citenamefont {Huang},
  \citenamefont {Zhu},\ and\ \citenamefont {Siew}}]{huang2006}%
  \BibitemOpen
  \bibfield  {author} {\bibinfo {author} {\bibfnamefont {Guang-Bin}\
  \bibnamefont {Huang}}, \bibinfo {author} {\bibfnamefont {Qin-Yu}\
  \bibnamefont {Zhu}}, \ and\ \bibinfo {author} {\bibfnamefont {Chee-Kheong}\
  \bibnamefont {Siew}},\ }\bibfield  {title} {\enquote {\bibinfo {title}
  {Extreme learning machine: {{Theory}} and applications},}\ }\href {\doibase
  10.1016/j.neucom.2005.12.126} {\bibfield  {journal} {\bibinfo  {journal}
  {Neurocomputing}\ }\bibinfo {series} {Neural {{Networks}}},\ \textbf
  {\bibinfo {volume} {70}},\ \bibinfo {pages} {489--501} (\bibinfo {year}
  {2006})}\BibitemShut {NoStop}%
\bibitem [{\citenamefont {Opala}\ \emph {et~al.}(2019)\citenamefont {Opala},
  \citenamefont {Ghosh}, \citenamefont {Liew},\ and\ \citenamefont
  {Matuszewski}}]{opala2019}%
  \BibitemOpen
  \bibfield  {author} {\bibinfo {author} {\bibfnamefont {Andrzej}\ \bibnamefont
  {Opala}}, \bibinfo {author} {\bibfnamefont {Sanjib}\ \bibnamefont {Ghosh}},
  \bibinfo {author} {\bibfnamefont {Timothy~C.H.}\ \bibnamefont {Liew}}, \ and\
  \bibinfo {author} {\bibfnamefont {Micha{\l}}\ \bibnamefont {Matuszewski}},\
  }\bibfield  {title} {\enquote {\bibinfo {title} {Neuromorphic {{Computing}}
  in {{Ginzburg-Landau Polariton-Lattice Systems}}},}\ }\href {\doibase
  10.1103/PhysRevApplied.11.064029} {\bibfield  {journal} {\bibinfo  {journal}
  {Physical Review Applied}\ }\textbf {\bibinfo {volume} {11}},\ \bibinfo
  {pages} {064029} (\bibinfo {year} {2019})}\BibitemShut {NoStop}%
\bibitem [{\citenamefont {Denis}\ \emph {et~al.}(2022)\citenamefont {Denis},
  \citenamefont {Favero},\ and\ \citenamefont {Ciuti}}]{denis2022}%
  \BibitemOpen
  \bibfield  {author} {\bibinfo {author} {\bibfnamefont {Zakari}\ \bibnamefont
  {Denis}}, \bibinfo {author} {\bibfnamefont {Ivan}\ \bibnamefont {Favero}}, \
  and\ \bibinfo {author} {\bibfnamefont {Cristiano}\ \bibnamefont {Ciuti}},\
  }\bibfield  {title} {\enquote {\bibinfo {title} {Photonic {{Kernel Machine
  Learning}} for {{Ultrafast Spectral Analysis}}},}\ }\href {\doibase
  10.1103/PhysRevApplied.17.034077} {\bibfield  {journal} {\bibinfo  {journal}
  {Physical Review Applied}\ }\textbf {\bibinfo {volume} {17}},\ \bibinfo
  {pages} {034077} (\bibinfo {year} {2022})}\BibitemShut {NoStop}%
\bibitem [{\citenamefont {Ballarini}\ \emph {et~al.}(2020)\citenamefont
  {Ballarini}, \citenamefont {Gianfrate}, \citenamefont {Panico}, \citenamefont
  {Opala}, \citenamefont {Ghosh}, \citenamefont {Dominici}, \citenamefont
  {Ardizzone}, \citenamefont {De~Giorgi}, \citenamefont {Lerario},
  \citenamefont {Gigli}, \citenamefont {Liew}, \citenamefont {Matuszewski},\
  and\ \citenamefont {Sanvitto}}]{ballarini2020}%
  \BibitemOpen
  \bibfield  {author} {\bibinfo {author} {\bibfnamefont {Dario}\ \bibnamefont
  {Ballarini}}, \bibinfo {author} {\bibfnamefont {Antonio}\ \bibnamefont
  {Gianfrate}}, \bibinfo {author} {\bibfnamefont {Riccardo}\ \bibnamefont
  {Panico}}, \bibinfo {author} {\bibfnamefont {Andrzej}\ \bibnamefont {Opala}},
  \bibinfo {author} {\bibfnamefont {Sanjib}\ \bibnamefont {Ghosh}}, \bibinfo
  {author} {\bibfnamefont {Lorenzo}\ \bibnamefont {Dominici}}, \bibinfo
  {author} {\bibfnamefont {Vincenzo}\ \bibnamefont {Ardizzone}}, \bibinfo
  {author} {\bibfnamefont {Milena}\ \bibnamefont {De~Giorgi}}, \bibinfo
  {author} {\bibfnamefont {Giovanni}\ \bibnamefont {Lerario}}, \bibinfo
  {author} {\bibfnamefont {Giuseppe}\ \bibnamefont {Gigli}}, \bibinfo {author}
  {\bibfnamefont {Timothy C.~H.}\ \bibnamefont {Liew}}, \bibinfo {author}
  {\bibfnamefont {Michal}\ \bibnamefont {Matuszewski}}, \ and\ \bibinfo
  {author} {\bibfnamefont {Daniele}\ \bibnamefont {Sanvitto}},\ }\bibfield
  {title} {\enquote {\bibinfo {title} {Polaritonic {{Neuromorphic Computing
  Outperforms Linear Classifiers}}},}\ }\href {\doibase
  10.1021/acs.nanolett.0c00435} {\bibfield  {journal} {\bibinfo  {journal}
  {Nano Letters}\ }\textbf {\bibinfo {volume} {20}},\ \bibinfo {pages}
  {3506--3512} (\bibinfo {year} {2020})}\BibitemShut {NoStop}%
\bibitem [{\citenamefont {Pierangeli}\ \emph {et~al.}(2021)\citenamefont
  {Pierangeli}, \citenamefont {Marcucci},\ and\ \citenamefont
  {Conti}}]{pierangeli21}%
  \BibitemOpen
  \bibfield  {author} {\bibinfo {author} {\bibfnamefont {Davide}\ \bibnamefont
  {Pierangeli}}, \bibinfo {author} {\bibfnamefont {Giulia}\ \bibnamefont
  {Marcucci}}, \ and\ \bibinfo {author} {\bibfnamefont {Claudio}\ \bibnamefont
  {Conti}},\ }\bibfield  {title} {\enquote {\bibinfo {title} {Photonic extreme
  learning machine by free-space optical propagation},}\ }\href {\doibase
  10.1364/PRJ.423531} {\bibfield  {journal} {\bibinfo  {journal} {Photon.
  Res.}\ }\textbf {\bibinfo {volume} {9}},\ \bibinfo {pages} {1446--1454}
  (\bibinfo {year} {2021})}\BibitemShut {NoStop}%
\bibitem [{\citenamefont {Hofmann}\ \emph {et~al.}(2008)\citenamefont
  {Hofmann}, \citenamefont {Sch{\"o}lkopf},\ and\ \citenamefont
  {Smola}}]{hofmann2008}%
  \BibitemOpen
  \bibfield  {author} {\bibinfo {author} {\bibfnamefont {Thomas}\ \bibnamefont
  {Hofmann}}, \bibinfo {author} {\bibfnamefont {Bernhard}\ \bibnamefont
  {Sch{\"o}lkopf}}, \ and\ \bibinfo {author} {\bibfnamefont {Alexander~J.}\
  \bibnamefont {Smola}},\ }\bibfield  {title} {\enquote {\bibinfo {title}
  {Kernel methods in machine learning},}\ }\href {\doibase
  10.1214/009053607000000677} {\bibfield  {journal} {\bibinfo  {journal} {The
  Annals of Statistics}\ }\textbf {\bibinfo {volume} {36}},\ \bibinfo {pages}
  {1171--1220} (\bibinfo {year} {2008})}\BibitemShut {NoStop}%
\bibitem [{\citenamefont {Jacot}\ \emph {et~al.}(2018)\citenamefont {Jacot},
  \citenamefont {Gabriel},\ and\ \citenamefont {Hongler}}]{jacot2018}%
  \BibitemOpen
  \bibfield  {author} {\bibinfo {author} {\bibfnamefont {Arthur}\ \bibnamefont
  {Jacot}}, \bibinfo {author} {\bibfnamefont {Franck}\ \bibnamefont {Gabriel}},
  \ and\ \bibinfo {author} {\bibfnamefont {Clement}\ \bibnamefont {Hongler}},\
  }\bibfield  {title} {\enquote {\bibinfo {title} {Neural {{Tangent Kernel}}:
  {{Convergence}} and {{Generalization}} in {{Neural Networks}}},}\ }in\
  \href@noop {} {\emph {\bibinfo {booktitle} {Advances in {{Neural Information
  Processing Systems}}}}},\ Vol.~\bibinfo {volume} {31}\ (\bibinfo  {publisher}
  {{Curran Associates, Inc.}},\ \bibinfo {year} {2018})\BibitemShut {NoStop}%
\bibitem [{\citenamefont {Biamonte}\ \emph {et~al.}(2017)\citenamefont
  {Biamonte}, \citenamefont {Wittek}, \citenamefont {Pancotti}, \citenamefont
  {Rebentrost}, \citenamefont {Wiebe},\ and\ \citenamefont
  {Lloyd}}]{biamonte2017}%
  \BibitemOpen
  \bibfield  {author} {\bibinfo {author} {\bibfnamefont {Jacob}\ \bibnamefont
  {Biamonte}}, \bibinfo {author} {\bibfnamefont {Peter}\ \bibnamefont
  {Wittek}}, \bibinfo {author} {\bibfnamefont {Nicola}\ \bibnamefont
  {Pancotti}}, \bibinfo {author} {\bibfnamefont {Patrick}\ \bibnamefont
  {Rebentrost}}, \bibinfo {author} {\bibfnamefont {Nathan}\ \bibnamefont
  {Wiebe}}, \ and\ \bibinfo {author} {\bibfnamefont {Seth}\ \bibnamefont
  {Lloyd}},\ }\bibfield  {title} {\enquote {\bibinfo {title} {Quantum machine
  learning},}\ }\href {\doibase 10.1038/nature23474} {\bibfield  {journal}
  {\bibinfo  {journal} {Nature}\ }\textbf {\bibinfo {volume} {549}},\ \bibinfo
  {pages} {195--202} (\bibinfo {year} {2017})}\BibitemShut {NoStop}%
\bibitem [{\citenamefont {Dunjko}\ and\ \citenamefont
  {Briegel}(2018)}]{dunjko2018}%
  \BibitemOpen
  \bibfield  {author} {\bibinfo {author} {\bibfnamefont {Vedran}\ \bibnamefont
  {Dunjko}}\ and\ \bibinfo {author} {\bibfnamefont {Hans~J.}\ \bibnamefont
  {Briegel}},\ }\bibfield  {title} {\enquote {\bibinfo {title} {Machine
  learning \& artificial intelligence in the quantum domain: A review of recent
  progress},}\ }\href {\doibase 10.1088/1361-6633/aab406} {\bibfield  {journal}
  {\bibinfo  {journal} {Reports on Progress in Physics}\ }\textbf {\bibinfo
  {volume} {81}},\ \bibinfo {pages} {074001} (\bibinfo {year}
  {2018})}\BibitemShut {NoStop}%
\bibitem [{\citenamefont {Schuld}\ \emph {et~al.}(2014)\citenamefont {Schuld},
  \citenamefont {Sinayskiy},\ and\ \citenamefont {Petruccione}}]{schuld2014}%
  \BibitemOpen
  \bibfield  {author} {\bibinfo {author} {\bibfnamefont {Maria}\ \bibnamefont
  {Schuld}}, \bibinfo {author} {\bibfnamefont {Ilya}\ \bibnamefont
  {Sinayskiy}}, \ and\ \bibinfo {author} {\bibfnamefont {Francesco}\
  \bibnamefont {Petruccione}},\ }\bibfield  {title} {\enquote {\bibinfo {title}
  {The quest for a {{Quantum Neural Network}}},}\ }\href {\doibase
  10.1007/s11128-014-0809-8} {\bibfield  {journal} {\bibinfo  {journal}
  {Quantum Information Processing}\ }\textbf {\bibinfo {volume} {13}},\
  \bibinfo {pages} {2567--2586} (\bibinfo {year} {2014})}\BibitemShut {NoStop}%
\bibitem [{\citenamefont {Markovi{\'c}}\ and\ \citenamefont
  {Grollier}(2020)}]{markovic2020}%
  \BibitemOpen
  \bibfield  {author} {\bibinfo {author} {\bibfnamefont {Danijela}\
  \bibnamefont {Markovi{\'c}}}\ and\ \bibinfo {author} {\bibfnamefont {Julie}\
  \bibnamefont {Grollier}},\ }\bibfield  {title} {\enquote {\bibinfo {title}
  {Quantum neuromorphic computing},}\ }\href {\doibase 10.1063/5.0020014}
  {\bibfield  {journal} {\bibinfo  {journal} {Applied Physics Letters}\
  }\textbf {\bibinfo {volume} {117}},\ \bibinfo {pages} {150501} (\bibinfo
  {year} {2020})}\BibitemShut {NoStop}%
\bibitem [{\citenamefont {Benedetti}\ \emph {et~al.}(2019)\citenamefont
  {Benedetti}, \citenamefont {Lloyd}, \citenamefont {Sack},\ and\ \citenamefont
  {Fiorentini}}]{benedetti2019}%
  \BibitemOpen
  \bibfield  {author} {\bibinfo {author} {\bibfnamefont {Marcello}\
  \bibnamefont {Benedetti}}, \bibinfo {author} {\bibfnamefont {Erika}\
  \bibnamefont {Lloyd}}, \bibinfo {author} {\bibfnamefont {Stefan}\
  \bibnamefont {Sack}}, \ and\ \bibinfo {author} {\bibfnamefont {Mattia}\
  \bibnamefont {Fiorentini}},\ }\bibfield  {title} {\enquote {\bibinfo {title}
  {Parameterized quantum circuits as machine learning models},}\ }\href
  {\doibase 10.1088/2058-9565/ab4eb5} {\bibfield  {journal} {\bibinfo
  {journal} {Quantum Science and Technology}\ }\textbf {\bibinfo {volume}
  {4}},\ \bibinfo {pages} {043001} (\bibinfo {year} {2019})}\BibitemShut
  {NoStop}%
\bibitem [{\citenamefont {Farhi}\ and\ \citenamefont
  {Neven}(2018)}]{farhi2018a}%
  \BibitemOpen
  \bibfield  {author} {\bibinfo {author} {\bibfnamefont {Edward}\ \bibnamefont
  {Farhi}}\ and\ \bibinfo {author} {\bibfnamefont {Hartmut}\ \bibnamefont
  {Neven}},\ }\bibfield  {title} {\enquote {\bibinfo {title} {Classification
  with {{Quantum Neural Networks}} on {{Near Term Processors}}},}\ }\href@noop
  {} {\  (\bibinfo {year} {2018})},\ \Eprint {http://arxiv.org/abs/1802.06002}
  {arXiv:1802.06002} \BibitemShut {NoStop}%
\bibitem [{\citenamefont {Park}\ \emph {et~al.}(2020)\citenamefont {Park},
  \citenamefont {Blank},\ and\ \citenamefont {Petruccione}}]{park2020}%
  \BibitemOpen
  \bibfield  {author} {\bibinfo {author} {\bibfnamefont {Daniel~K.}\
  \bibnamefont {Park}}, \bibinfo {author} {\bibfnamefont {Carsten}\
  \bibnamefont {Blank}}, \ and\ \bibinfo {author} {\bibfnamefont {Francesco}\
  \bibnamefont {Petruccione}},\ }\bibfield  {title} {\enquote {\bibinfo {title}
  {The theory of the quantum kernel-based binary classifier},}\ }\href
  {\doibase 10.1016/j.physleta.2020.126422} {\bibfield  {journal} {\bibinfo
  {journal} {Physics Letters A}\ }\textbf {\bibinfo {volume} {384}},\ \bibinfo
  {pages} {126422} (\bibinfo {year} {2020})}\BibitemShut {NoStop}%
\bibitem [{\citenamefont {Schuld}\ and\ \citenamefont
  {Killoran}(2019)}]{schuld2019}%
  \BibitemOpen
  \bibfield  {author} {\bibinfo {author} {\bibfnamefont {Maria}\ \bibnamefont
  {Schuld}}\ and\ \bibinfo {author} {\bibfnamefont {Nathan}\ \bibnamefont
  {Killoran}},\ }\bibfield  {title} {\enquote {\bibinfo {title} {Quantum
  {{Machine Learning}} in {{Feature Hilbert Spaces}}},}\ }\href {\doibase
  10.1103/PhysRevLett.122.040504} {\bibfield  {journal} {\bibinfo  {journal}
  {Physical Review Letters}\ }\textbf {\bibinfo {volume} {122}},\ \bibinfo
  {pages} {040504} (\bibinfo {year} {2019})}\BibitemShut {NoStop}%
\bibitem [{\citenamefont {Lloyd}\ \emph {et~al.}(2020)\citenamefont {Lloyd},
  \citenamefont {Schuld}, \citenamefont {Ijaz}, \citenamefont {Izaac},\ and\
  \citenamefont {Killoran}}]{lloyd2020}%
  \BibitemOpen
  \bibfield  {author} {\bibinfo {author} {\bibfnamefont {Seth}\ \bibnamefont
  {Lloyd}}, \bibinfo {author} {\bibfnamefont {Maria}\ \bibnamefont {Schuld}},
  \bibinfo {author} {\bibfnamefont {Aroosa}\ \bibnamefont {Ijaz}}, \bibinfo
  {author} {\bibfnamefont {Josh}\ \bibnamefont {Izaac}}, \ and\ \bibinfo
  {author} {\bibfnamefont {Nathan}\ \bibnamefont {Killoran}},\ }\bibfield
  {title} {\enquote {\bibinfo {title} {Quantum embeddings for machine
  learning},}\ }\href {\doibase 10.48550/arXiv.2001.03622} {\  (\bibinfo {year}
  {2020}),\ 10.48550/arXiv.2001.03622}\BibitemShut {NoStop}%
\bibitem [{\citenamefont {Hubregtsen}\ \emph {et~al.}(2021)\citenamefont
  {Hubregtsen}, \citenamefont {Wierichs}, \citenamefont {{Gil-Fuster}},
  \citenamefont {Derks}, \citenamefont {Faehrmann},\ and\ \citenamefont
  {Meyer}}]{hubregtsen2021}%
  \BibitemOpen
  \bibfield  {author} {\bibinfo {author} {\bibfnamefont {Thomas}\ \bibnamefont
  {Hubregtsen}}, \bibinfo {author} {\bibfnamefont {David}\ \bibnamefont
  {Wierichs}}, \bibinfo {author} {\bibfnamefont {Elies}\ \bibnamefont
  {{Gil-Fuster}}}, \bibinfo {author} {\bibfnamefont {Peter-Jan H.~S.}\
  \bibnamefont {Derks}}, \bibinfo {author} {\bibfnamefont {Paul~K.}\
  \bibnamefont {Faehrmann}}, \ and\ \bibinfo {author} {\bibfnamefont
  {Johannes~Jakob}\ \bibnamefont {Meyer}},\ }\bibfield  {title} {\enquote
  {\bibinfo {title} {Training {{Quantum Embedding Kernels}} on {{Near-Term
  Quantum Computers}}},}\ }\href@noop {} {\  (\bibinfo {year} {2021})},\
  \Eprint {http://arxiv.org/abs/2105.02276} {arXiv:2105.02276} \BibitemShut
  {NoStop}%
\bibitem [{\citenamefont {Schuld}(2021)}]{schuld2021}%
  \BibitemOpen
  \bibfield  {author} {\bibinfo {author} {\bibfnamefont {Maria}\ \bibnamefont
  {Schuld}},\ }\bibfield  {title} {\enquote {\bibinfo {title} {Supervised
  quantum machine learning models are kernel methods},}\ }\href@noop {} {\
  (\bibinfo {year} {2021})},\ \Eprint {http://arxiv.org/abs/2101.11020}
  {arXiv:2101.11020} \BibitemShut {NoStop}%
\bibitem [{\citenamefont {Kusumoto}\ \emph {et~al.}(2021)\citenamefont
  {Kusumoto}, \citenamefont {Mitarai}, \citenamefont {Fujii}, \citenamefont
  {Kitagawa},\ and\ \citenamefont {Negoro}}]{kusumoto2021}%
  \BibitemOpen
  \bibfield  {author} {\bibinfo {author} {\bibfnamefont {Takeru}\ \bibnamefont
  {Kusumoto}}, \bibinfo {author} {\bibfnamefont {Kosuke}\ \bibnamefont
  {Mitarai}}, \bibinfo {author} {\bibfnamefont {Keisuke}\ \bibnamefont
  {Fujii}}, \bibinfo {author} {\bibfnamefont {Masahiro}\ \bibnamefont
  {Kitagawa}}, \ and\ \bibinfo {author} {\bibfnamefont {Makoto}\ \bibnamefont
  {Negoro}},\ }\bibfield  {title} {\enquote {\bibinfo {title} {Experimental
  quantum kernel trick with nuclear spins in a solid},}\ }\href {\doibase
  10.1038/s41534-021-00423-0} {\bibfield  {journal} {\bibinfo  {journal} {npj
  Quantum Information}\ }\textbf {\bibinfo {volume} {7}},\ \bibinfo {pages}
  {1--7} (\bibinfo {year} {2021})}\BibitemShut {NoStop}%
\bibitem [{\citenamefont {Liu}\ \emph {et~al.}(2021)\citenamefont {Liu},
  \citenamefont {Arunachalam},\ and\ \citenamefont {Temme}}]{liu2021}%
  \BibitemOpen
  \bibfield  {author} {\bibinfo {author} {\bibfnamefont {Yunchao}\ \bibnamefont
  {Liu}}, \bibinfo {author} {\bibfnamefont {Srinivasan}\ \bibnamefont
  {Arunachalam}}, \ and\ \bibinfo {author} {\bibfnamefont {Kristan}\
  \bibnamefont {Temme}},\ }\bibfield  {title} {\enquote {\bibinfo {title} {A
  rigorous and robust quantum speed-up in supervised machine learning},}\
  }\href {\doibase 10.1038/s41567-021-01287-z} {\bibfield  {journal} {\bibinfo
  {journal} {Nature Physics}\ }\textbf {\bibinfo {volume} {17}},\ \bibinfo
  {pages} {1013--1017} (\bibinfo {year} {2021})}\BibitemShut {NoStop}%
\bibitem [{\citenamefont {Wu}\ \emph {et~al.}(2021)\citenamefont {Wu},
  \citenamefont {Wu}, \citenamefont {Wang},\ and\ \citenamefont
  {Yuan}}]{wu2021}%
  \BibitemOpen
  \bibfield  {author} {\bibinfo {author} {\bibfnamefont {Yusen}\ \bibnamefont
  {Wu}}, \bibinfo {author} {\bibfnamefont {Bujiao}\ \bibnamefont {Wu}},
  \bibinfo {author} {\bibfnamefont {Jingbo}\ \bibnamefont {Wang}}, \ and\
  \bibinfo {author} {\bibfnamefont {Xiao}\ \bibnamefont {Yuan}},\ }\bibfield
  {title} {\enquote {\bibinfo {title} {Provable {{Advantage}} in {{Quantum
  Phase Learning}} via {{Quantum Kernel Alphatron}}},}\ }\href@noop {} {\
  (\bibinfo {year} {2021})},\ \Eprint {http://arxiv.org/abs/2111.07553}
  {arXiv:2111.07553} \BibitemShut {NoStop}%
\bibitem [{\citenamefont {K{\"u}bler}\ \emph {et~al.}(2021)\citenamefont
  {K{\"u}bler}, \citenamefont {Buchholz},\ and\ \citenamefont
  {Sch{\"o}lkopf}}]{kubler2021}%
  \BibitemOpen
  \bibfield  {author} {\bibinfo {author} {\bibfnamefont {Jonas~M.}\
  \bibnamefont {K{\"u}bler}}, \bibinfo {author} {\bibfnamefont {Simon}\
  \bibnamefont {Buchholz}}, \ and\ \bibinfo {author} {\bibfnamefont {Bernhard}\
  \bibnamefont {Sch{\"o}lkopf}},\ }\bibfield  {title} {\enquote {\bibinfo
  {title} {The {{Inductive Bias}} of {{Quantum Kernels}}},}\ }\href {\doibase
  10.48550/arXiv.2106.03747} {\  (\bibinfo {year} {2021}),\
  10.48550/arXiv.2106.03747}\BibitemShut {NoStop}%
\bibitem [{\citenamefont {Shaydulin}\ and\ \citenamefont
  {Wild}(2021)}]{shaydulin2021}%
  \BibitemOpen
  \bibfield  {author} {\bibinfo {author} {\bibfnamefont {Ruslan}\ \bibnamefont
  {Shaydulin}}\ and\ \bibinfo {author} {\bibfnamefont {Stefan~M.}\ \bibnamefont
  {Wild}},\ }\bibfield  {title} {\enquote {\bibinfo {title} {Importance of
  {{Kernel Bandwidth}} in {{Quantum Machine Learning}}},}\ }\href@noop {} {\
  (\bibinfo {year} {2021})},\ \Eprint {http://arxiv.org/abs/2111.05451}
  {arXiv:2111.05451} \BibitemShut {NoStop}%
\bibitem [{\citenamefont {Wang}\ \emph {et~al.}(2021)\citenamefont {Wang},
  \citenamefont {Du}, \citenamefont {Luo},\ and\ \citenamefont
  {Tao}}]{wang2021}%
  \BibitemOpen
  \bibfield  {author} {\bibinfo {author} {\bibfnamefont {Xinbiao}\ \bibnamefont
  {Wang}}, \bibinfo {author} {\bibfnamefont {Yuxuan}\ \bibnamefont {Du}},
  \bibinfo {author} {\bibfnamefont {Yong}\ \bibnamefont {Luo}}, \ and\ \bibinfo
  {author} {\bibfnamefont {Dacheng}\ \bibnamefont {Tao}},\ }\bibfield  {title}
  {\enquote {\bibinfo {title} {Towards understanding the power of quantum
  kernels in the {{NISQ}} era},}\ }\href {\doibase 10.22331/q-2021-08-30-531}
  {\bibfield  {journal} {\bibinfo  {journal} {Quantum}\ }\textbf {\bibinfo
  {volume} {5}},\ \bibinfo {pages} {531} (\bibinfo {year} {2021})}\BibitemShut
  {NoStop}%
\bibitem [{\citenamefont {Bartkiewicz}\ \emph {et~al.}(2020)\citenamefont
  {Bartkiewicz}, \citenamefont {Gneiting}, \citenamefont {{\v C}ernoch},
  \citenamefont {Jir{\'a}kov{\'a}}, \citenamefont {Lemr},\ and\ \citenamefont
  {Nori}}]{bartkiewicz2020}%
  \BibitemOpen
  \bibfield  {author} {\bibinfo {author} {\bibfnamefont {Karol}\ \bibnamefont
  {Bartkiewicz}}, \bibinfo {author} {\bibfnamefont {Clemens}\ \bibnamefont
  {Gneiting}}, \bibinfo {author} {\bibfnamefont {Anton{\'i}n}\ \bibnamefont
  {{\v C}ernoch}}, \bibinfo {author} {\bibfnamefont {Kate{\v r}ina}\
  \bibnamefont {Jir{\'a}kov{\'a}}}, \bibinfo {author} {\bibfnamefont {Karel}\
  \bibnamefont {Lemr}}, \ and\ \bibinfo {author} {\bibfnamefont {Franco}\
  \bibnamefont {Nori}},\ }\bibfield  {title} {\enquote {\bibinfo {title}
  {Experimental kernel-based quantum machine learning in finite feature
  space},}\ }\href {\doibase 10.1038/s41598-020-68911-5} {\bibfield  {journal}
  {\bibinfo  {journal} {Scientific Reports}\ }\textbf {\bibinfo {volume}
  {10}},\ \bibinfo {pages} {12356} (\bibinfo {year} {2020})}\BibitemShut
  {NoStop}%
\bibitem [{\citenamefont {Preskill}(2018)}]{preskill2018}%
  \BibitemOpen
  \bibfield  {author} {\bibinfo {author} {\bibfnamefont {John}\ \bibnamefont
  {Preskill}},\ }\bibfield  {title} {\enquote {\bibinfo {title} {Quantum
  {{Computing}} in the {{NISQ}} era and beyond},}\ }\href {\doibase
  10.22331/q-2018-08-06-79} {\bibfield  {journal} {\bibinfo  {journal}
  {Quantum}\ }\textbf {\bibinfo {volume} {2}},\ \bibinfo {pages} {79} (\bibinfo
  {year} {2018})}\BibitemShut {NoStop}%
\bibitem [{\citenamefont {Fujii}\ and\ \citenamefont
  {Nakajima}(2017)}]{fujii2017}%
  \BibitemOpen
  \bibfield  {author} {\bibinfo {author} {\bibfnamefont {Keisuke}\ \bibnamefont
  {Fujii}}\ and\ \bibinfo {author} {\bibfnamefont {Kohei}\ \bibnamefont
  {Nakajima}},\ }\bibfield  {title} {\enquote {\bibinfo {title} {Harnessing
  {{Disordered-Ensemble Quantum Dynamics}} for {{Machine Learning}}},}\ }\href
  {\doibase 10.1103/PhysRevApplied.8.024030} {\bibfield  {journal} {\bibinfo
  {journal} {Physical Review Applied}\ }\textbf {\bibinfo {volume} {8}},\
  \bibinfo {pages} {024030} (\bibinfo {year} {2017})}\BibitemShut {NoStop}%
\bibitem [{\citenamefont {Xu}\ \emph {et~al.}(2021)\citenamefont {Xu},
  \citenamefont {Krisnanda}, \citenamefont {Verstraelen}, \citenamefont
  {Liew},\ and\ \citenamefont {Ghosh}}]{xu2021}%
  \BibitemOpen
  \bibfield  {author} {\bibinfo {author} {\bibfnamefont {Huawen}\ \bibnamefont
  {Xu}}, \bibinfo {author} {\bibfnamefont {Tanjung}\ \bibnamefont {Krisnanda}},
  \bibinfo {author} {\bibfnamefont {Wouter}\ \bibnamefont {Verstraelen}},
  \bibinfo {author} {\bibfnamefont {Timothy C.~H.}\ \bibnamefont {Liew}}, \
  and\ \bibinfo {author} {\bibfnamefont {Sanjib}\ \bibnamefont {Ghosh}},\
  }\bibfield  {title} {\enquote {\bibinfo {title} {Superpolynomial quantum
  enhancement in polaritonic neuromorphic computing},}\ }\href {\doibase
  10.1103/PhysRevB.103.195302} {\bibfield  {journal} {\bibinfo  {journal}
  {Physical Review B}\ }\textbf {\bibinfo {volume} {103}},\ \bibinfo {pages}
  {195302} (\bibinfo {year} {2021})}\BibitemShut {NoStop}%
\bibitem [{Note1()}]{Note1}%
  \BibitemOpen
  \bibinfo {note} {In the case of multivariate functions the variables against
  which the expectation is taken is indicated in subscript.}\BibitemShut
  {Stop}%
\bibitem [{\citenamefont {Breuer}\ and\ \citenamefont
  {Petruccione}(2007)}]{breuer2007}%
  \BibitemOpen
  \bibfield  {author} {\bibinfo {author} {\bibfnamefont {Heinz-Peter}\
  \bibnamefont {Breuer}}\ and\ \bibinfo {author} {\bibfnamefont {Francesco}\
  \bibnamefont {Petruccione}},\ }\href {\doibase
  10.1093/acprof:oso/9780199213900.001.0001} {\emph {\bibinfo {title} {The
  {{Theory}} of {{Open Quantum Systems}}}}}\ (\bibinfo  {publisher} {{Oxford
  University Press}},\ \bibinfo {address} {{Oxford}},\ \bibinfo {year}
  {2007})\BibitemShut {NoStop}%
\bibitem [{\citenamefont {Suykens}\ and\ \citenamefont
  {Vandewalle}(1999)}]{suykens1999}%
  \BibitemOpen
  \bibfield  {author} {\bibinfo {author} {\bibfnamefont {J.A.K.}\ \bibnamefont
  {Suykens}}\ and\ \bibinfo {author} {\bibfnamefont {J.}~\bibnamefont
  {Vandewalle}},\ }\bibfield  {title} {\enquote {\bibinfo {title} {Least
  {{Squares Support Vector Machine Classifiers}}},}\ }\href {\doibase
  10.1023/A:1018628609742} {\bibfield  {journal} {\bibinfo  {journal} {Neural
  Processing Letters}\ }\textbf {\bibinfo {volume} {9}},\ \bibinfo {pages}
  {293--300} (\bibinfo {year} {1999})}\BibitemShut {NoStop}%
\bibitem [{\citenamefont {Hastie}\ \emph {et~al.}(2013)\citenamefont {Hastie},
  \citenamefont {Tibshirani},\ and\ \citenamefont {Friedman}}]{hastie2013}%
  \BibitemOpen
  \bibfield  {author} {\bibinfo {author} {\bibfnamefont {Trevor}\ \bibnamefont
  {Hastie}}, \bibinfo {author} {\bibfnamefont {Robert}\ \bibnamefont
  {Tibshirani}}, \ and\ \bibinfo {author} {\bibfnamefont {Jerome}\ \bibnamefont
  {Friedman}},\ }\href@noop {} {\emph {\bibinfo {title} {The {{Elements}} of
  {{Statistical Learning}}: {{Data Mining}}, {{Inference}}, and
  {{Prediction}}}}}\ (\bibinfo  {publisher} {{Springer Science \& Business
  Media}},\ \bibinfo {year} {2013})\BibitemShut {NoStop}%
\bibitem [{Note2()}]{Note2}%
  \BibitemOpen
  \bibinfo {note} {Note that in certain scenario, one may prefer to exclude the
  first component of $\protect \mathaccentV {vec}17E{w}$, which is a constant
  intercept term, in the regularization. Then it suffices to replace the
  $(1,1)$ entry of $\protect \mathds {1}$ by $0$ in Eq.~\protect \textup {\hbox
  {\mathsurround \z@ \protect \normalfont (\ignorespaces \ref
  {eq:optimal_weights}\unskip \@@italiccorr )}} to obtain the optimal weights.
  This is equivalent to using a centered kernel, as discussed in Appendix \ref
  {appendix:B_intercept_and_kernel_centering}.}\BibitemShut {Stop}%
\bibitem [{\citenamefont {Jerbi}\ \emph {et~al.}(2021)\citenamefont {Jerbi},
  \citenamefont {Fiderer}, \citenamefont {Nautrup}, \citenamefont {K{\"u}bler},
  \citenamefont {Briegel},\ and\ \citenamefont {Dunjko}}]{jerbi2021}%
  \BibitemOpen
  \bibfield  {author} {\bibinfo {author} {\bibfnamefont {Sofiene}\ \bibnamefont
  {Jerbi}}, \bibinfo {author} {\bibfnamefont {Lukas~J.}\ \bibnamefont
  {Fiderer}}, \bibinfo {author} {\bibfnamefont {Hendrik~Poulsen}\ \bibnamefont
  {Nautrup}}, \bibinfo {author} {\bibfnamefont {Jonas~M.}\ \bibnamefont
  {K{\"u}bler}}, \bibinfo {author} {\bibfnamefont {Hans~J.}\ \bibnamefont
  {Briegel}}, \ and\ \bibinfo {author} {\bibfnamefont {Vedran}\ \bibnamefont
  {Dunjko}},\ }\bibfield  {title} {\enquote {\bibinfo {title} {Quantum machine
  learning beyond kernel methods},}\ }\href@noop {} {\  (\bibinfo {year}
  {2021})},\ \Eprint {http://arxiv.org/abs/2110.13162} {arXiv:2110.13162}
  \BibitemShut {NoStop}%
\bibitem [{Note3()}]{Note3}%
  \BibitemOpen
  \bibinfo {note} {The probability measure $p$ on the input space $\protect
  \mathcal {X}$ is important here as it determines the scalar product on the
  space of real-valued functions on $\protect \mathcal {X}$ through
  {$\delimiter "426830A f,g\delimiter "526930B = \protect \mathbb {E}_{\vb
  *{x}\sim p}\left [ f(\vb *{x})g(\vb *{x})\right ]$}. The reproducing
  property, crucial to link the RKHS and its kernel, relies on such a
  well-defined scalar product.}\BibitemShut {Stop}%
\bibitem [{\citenamefont {Paulsen}\ and\ \citenamefont
  {Raghupathi}(2016)}]{paulsen2016}%
  \BibitemOpen
  \bibfield  {author} {\bibinfo {author} {\bibfnamefont {Vern~I.}\ \bibnamefont
  {Paulsen}}\ and\ \bibinfo {author} {\bibfnamefont {Mrinal}\ \bibnamefont
  {Raghupathi}},\ }\href {\doibase 10.1017/CBO9781316219232} {\emph {\bibinfo
  {title} {An {{Introduction}} to the {{Theory}} of {{Reproducing Kernel
  Hilbert Spaces}}}}},\ Cambridge {{Studies}} in {{Advanced Mathematics}}\
  (\bibinfo  {publisher} {{Cambridge University Press}},\ \bibinfo {address}
  {{Cambridge}},\ \bibinfo {year} {2016})\BibitemShut {NoStop}%
\bibitem [{\citenamefont {Schuld}\ \emph {et~al.}(2021)\citenamefont {Schuld},
  \citenamefont {Sweke},\ and\ \citenamefont {Meyer}}]{schuld2021a}%
  \BibitemOpen
  \bibfield  {author} {\bibinfo {author} {\bibfnamefont {Maria}\ \bibnamefont
  {Schuld}}, \bibinfo {author} {\bibfnamefont {Ryan}\ \bibnamefont {Sweke}}, \
  and\ \bibinfo {author} {\bibfnamefont {Johannes~Jakob}\ \bibnamefont
  {Meyer}},\ }\bibfield  {title} {\enquote {\bibinfo {title} {Effect of data
  encoding on the expressive power of variational quantum-machine-learning
  models},}\ }\href {\doibase 10.1103/PhysRevA.103.032430} {\bibfield
  {journal} {\bibinfo  {journal} {Physical Review A}\ }\textbf {\bibinfo
  {volume} {103}},\ \bibinfo {pages} {032430} (\bibinfo {year}
  {2021})}\BibitemShut {NoStop}%
\bibitem [{Note4()}]{Note4}%
  \BibitemOpen
  \bibinfo {note} {For a closed system, this quantum kernel can be directly
  evaluated through measurement \cite {park2020} and the trial function can be
  expressed in terms of the quantum kernel and optimized in an equivalent
  way.}\BibitemShut {Stop}%
\bibitem [{\citenamefont {Bordelon}\ \emph {et~al.}(2021)\citenamefont
  {Bordelon}, \citenamefont {Canatar},\ and\ \citenamefont
  {Pehlevan}}]{bordelon2021}%
  \BibitemOpen
  \bibfield  {author} {\bibinfo {author} {\bibfnamefont {Blake}\ \bibnamefont
  {Bordelon}}, \bibinfo {author} {\bibfnamefont {Abdulkadir}\ \bibnamefont
  {Canatar}}, \ and\ \bibinfo {author} {\bibfnamefont {Cengiz}\ \bibnamefont
  {Pehlevan}},\ }\bibfield  {title} {\enquote {\bibinfo {title} {Spectrum
  {{Dependent Learning Curves}} in {{Kernel Regression}} and {{Wide Neural
  Networks}}},}\ }\href@noop {} {\  (\bibinfo {year} {2021})},\ \Eprint
  {http://arxiv.org/abs/2002.02561} {arXiv:2002.02561} \BibitemShut {NoStop}%
\bibitem [{\citenamefont {Canatar}\ \emph {et~al.}(2021)\citenamefont
  {Canatar}, \citenamefont {Bordelon},\ and\ \citenamefont
  {Pehlevan}}]{canatar2021}%
  \BibitemOpen
  \bibfield  {author} {\bibinfo {author} {\bibfnamefont {Abdulkadir}\
  \bibnamefont {Canatar}}, \bibinfo {author} {\bibfnamefont {Blake}\
  \bibnamefont {Bordelon}}, \ and\ \bibinfo {author} {\bibfnamefont {Cengiz}\
  \bibnamefont {Pehlevan}},\ }\bibfield  {title} {\enquote {\bibinfo {title}
  {Spectral bias and task-model alignment explain generalization in kernel
  regression and infinitely wide neural networks},}\ }\href {\doibase
  10.1038/s41467-021-23103-1} {\bibfield  {journal} {\bibinfo  {journal}
  {Nature Communications}\ }\textbf {\bibinfo {volume} {12}},\ \bibinfo {pages}
  {2914} (\bibinfo {year} {2021})}\BibitemShut {NoStop}%
\bibitem [{\citenamefont {Abbas}\ \emph {et~al.}(2021)\citenamefont {Abbas},
  \citenamefont {Sutter}, \citenamefont {Zoufal}, \citenamefont {Lucchi},
  \citenamefont {Figalli},\ and\ \citenamefont {Woerner}}]{abbas2021}%
  \BibitemOpen
  \bibfield  {author} {\bibinfo {author} {\bibfnamefont {Amira}\ \bibnamefont
  {Abbas}}, \bibinfo {author} {\bibfnamefont {David}\ \bibnamefont {Sutter}},
  \bibinfo {author} {\bibfnamefont {Christa}\ \bibnamefont {Zoufal}}, \bibinfo
  {author} {\bibfnamefont {Aurelien}\ \bibnamefont {Lucchi}}, \bibinfo {author}
  {\bibfnamefont {Alessio}\ \bibnamefont {Figalli}}, \ and\ \bibinfo {author}
  {\bibfnamefont {Stefan}\ \bibnamefont {Woerner}},\ }\bibfield  {title}
  {\enquote {\bibinfo {title} {The power of quantum neural networks},}\ }\href
  {\doibase 10.1038/s43588-021-00084-1} {\bibfield  {journal} {\bibinfo
  {journal} {Nature Computational Science}\ }\textbf {\bibinfo {volume} {1}},\
  \bibinfo {pages} {403--409} (\bibinfo {year} {2021})}\BibitemShut {NoStop}%
\bibitem [{\citenamefont {Hastie}\ and\ \citenamefont
  {Zhu}(2006)}]{hastie2006}%
  \BibitemOpen
  \bibfield  {author} {\bibinfo {author} {\bibfnamefont {Trevor}\ \bibnamefont
  {Hastie}}\ and\ \bibinfo {author} {\bibfnamefont {Ji}~\bibnamefont {Zhu}},\
  }\bibfield  {title} {\enquote {\bibinfo {title} {Comment: [support vector
  machines with applications]},}\ }\href {http://www.jstor.org/stable/27645769}
  {\bibfield  {journal} {\bibinfo  {journal} {Statistical Science}\ }\textbf
  {\bibinfo {volume} {21}},\ \bibinfo {pages} {352--357} (\bibinfo {year}
  {2006})}\BibitemShut {NoStop}%
\bibitem [{\citenamefont {Huang}\ \emph {et~al.}(2021)\citenamefont {Huang},
  \citenamefont {Broughton}, \citenamefont {Mohseni}, \citenamefont {Babbush},
  \citenamefont {Boixo}, \citenamefont {Neven},\ and\ \citenamefont
  {McClean}}]{huang2021}%
  \BibitemOpen
  \bibfield  {author} {\bibinfo {author} {\bibfnamefont {Hsin-Yuan}\
  \bibnamefont {Huang}}, \bibinfo {author} {\bibfnamefont {Michael}\
  \bibnamefont {Broughton}}, \bibinfo {author} {\bibfnamefont {Masoud}\
  \bibnamefont {Mohseni}}, \bibinfo {author} {\bibfnamefont {Ryan}\
  \bibnamefont {Babbush}}, \bibinfo {author} {\bibfnamefont {Sergio}\
  \bibnamefont {Boixo}}, \bibinfo {author} {\bibfnamefont {Hartmut}\
  \bibnamefont {Neven}}, \ and\ \bibinfo {author} {\bibfnamefont {Jarrod~R.}\
  \bibnamefont {McClean}},\ }\bibfield  {title} {\enquote {\bibinfo {title}
  {Power of data in quantum machine learning},}\ }\href {\doibase
  10.1038/s41467-021-22539-9} {\bibfield  {journal} {\bibinfo  {journal}
  {Nature Communications}\ }\textbf {\bibinfo {volume} {12}},\ \bibinfo {pages}
  {2631} (\bibinfo {year} {2021})}\BibitemShut {NoStop}%
\bibitem [{\citenamefont {Cristianini}\ \emph {et~al.}(2006)\citenamefont
  {Cristianini}, \citenamefont {Kandola}, \citenamefont {Elisseeff},\ and\
  \citenamefont {{Shawe-Taylor}}}]{cristianini2006}%
  \BibitemOpen
  \bibfield  {author} {\bibinfo {author} {\bibfnamefont {Nello}\ \bibnamefont
  {Cristianini}}, \bibinfo {author} {\bibfnamefont {Jaz}\ \bibnamefont
  {Kandola}}, \bibinfo {author} {\bibfnamefont {Andre}\ \bibnamefont
  {Elisseeff}}, \ and\ \bibinfo {author} {\bibfnamefont {John}\ \bibnamefont
  {{Shawe-Taylor}}},\ }\bibfield  {title} {\enquote {\bibinfo {title} {On
  {{Kernel Target Alignment}}},}\ }in\ \href {\doibase 10.1007/3-540-33486-6_8}
  {\emph {\bibinfo {booktitle} {Innovations in {{Machine Learning}}: {{Theory}}
  and {{Applications}}}}},\ \bibinfo {series and number} {Studies in
  {{Fuzziness}} and {{Soft Computing}}},\ \bibinfo {editor} {edited by\
  \bibinfo {editor} {\bibfnamefont {Dawn~E.}\ \bibnamefont {Holmes}}\ and\
  \bibinfo {editor} {\bibfnamefont {Lakhmi~C.}\ \bibnamefont {Jain}}}\
  (\bibinfo  {publisher} {{Springer}},\ \bibinfo {address} {{Berlin,
  Heidelberg}},\ \bibinfo {year} {2006})\ pp.\ \bibinfo {pages}
  {205--256}\BibitemShut {NoStop}%
\bibitem [{\citenamefont {Mohri}\ \emph {et~al.}(2018)\citenamefont {Mohri},
  \citenamefont {Rostamizadeh},\ and\ \citenamefont
  {Talwalkar}}]{mohri2018foundations}%
  \BibitemOpen
  \bibfield  {author} {\bibinfo {author} {\bibfnamefont {Mehryar}\ \bibnamefont
  {Mohri}}, \bibinfo {author} {\bibfnamefont {Afshin}\ \bibnamefont
  {Rostamizadeh}}, \ and\ \bibinfo {author} {\bibfnamefont {Ameet}\
  \bibnamefont {Talwalkar}},\ }\href@noop {} {\emph {\bibinfo {title}
  {Foundations of machine learning}}}\ (\bibinfo  {publisher} {MIT press},\
  \bibinfo {year} {2018})\BibitemShut {NoStop}%
\bibitem [{\citenamefont {Banchi}\ \emph {et~al.}(2021)\citenamefont {Banchi},
  \citenamefont {Pereira},\ and\ \citenamefont {Pirandola}}]{banchi2021}%
  \BibitemOpen
  \bibfield  {author} {\bibinfo {author} {\bibfnamefont {Leonardo}\
  \bibnamefont {Banchi}}, \bibinfo {author} {\bibfnamefont {Jason}\
  \bibnamefont {Pereira}}, \ and\ \bibinfo {author} {\bibfnamefont {Stefano}\
  \bibnamefont {Pirandola}},\ }\bibfield  {title} {\enquote {\bibinfo {title}
  {Generalization in {{Quantum Machine Learning}}: {{A Quantum Information
  Standpoint}}},}\ }\href {\doibase 10.1103/PRXQuantum.2.040321} {\bibfield
  {journal} {\bibinfo  {journal} {PRX Quantum}\ }\textbf {\bibinfo {volume}
  {2}},\ \bibinfo {pages} {040321} (\bibinfo {year} {2021})}\BibitemShut
  {NoStop}%
\bibitem [{Note5()}]{Note5}%
  \BibitemOpen
  \bibinfo {note} {In total, the results presented in this work required
  approximately $800{,}000$ scalar hours ($\sim 90$ years) of computation and
  $5$ terabytes of storage on the acknowledged French National High Performance
  Computing facility (GENCI). During the simulations, we used approximately up
  to $30,000$ cores simultaneously.}\BibitemShut {Stop}%
\bibitem [{Note6()}]{Note6}%
  \BibitemOpen
  \bibinfo {note} {This can be explicitly expressed as a $D$-deep circuit via
  Trotterization as $\protect \mathcal {G} = \left [\DOTSB \prod@ \slimits@
  _{i=1}^N R_i^Z\(-\protect \genfrac {}{}{}1{2\Delta _i \Delta t}{D}\) \DOTSB
  \prod@ \slimits@ _{\delimiter "426830A i, j \delimiter "526930B }\DOTSB
  \prod@ \slimits@ _{K\in \protect \{X,Y,Z\protect \}} R_{ij}^{KK}(\protect
  \genfrac {}{}{}1{2 J_{ij}^K \Delta t}{D}) \right ]^D + O(J_0^2 \Delta
  t^2/D^2)$}\BibitemShut {NoStop}%
\bibitem [{\citenamefont {Czerwinski}(2021)}]{czerwinski2021}%
  \BibitemOpen
  \bibfield  {author} {\bibinfo {author} {\bibfnamefont {Artur}\ \bibnamefont
  {Czerwinski}},\ }\bibfield  {title} {\enquote {\bibinfo {title} {Quantum
  state tomography with informationally complete {{POVMs}} generated in the
  time domain},}\ }\href {\doibase 10.1007/s11128-021-03045-9} {\bibfield
  {journal} {\bibinfo  {journal} {Quantum Information Processing}\ }\textbf
  {\bibinfo {volume} {20}},\ \bibinfo {pages} {105} (\bibinfo {year}
  {2021})}\BibitemShut {NoStop}%
\bibitem [{\citenamefont {Di~Lorenzo}(2013)}]{dilorenzo2013}%
  \BibitemOpen
  \bibfield  {author} {\bibinfo {author} {\bibfnamefont {Antonio}\ \bibnamefont
  {Di~Lorenzo}},\ }\bibfield  {title} {\enquote {\bibinfo {title} {Sequential
  {{Measurement}} of {{Conjugate Variables}} as an {{Alternative Quantum State
  Tomography}}},}\ }\href {\doibase 10.1103/PhysRevLett.110.010404} {\bibfield
  {journal} {\bibinfo  {journal} {Physical Review Letters}\ }\textbf {\bibinfo
  {volume} {110}},\ \bibinfo {pages} {010404} (\bibinfo {year}
  {2013})}\BibitemShut {NoStop}%
\bibitem [{\citenamefont {Williams}\ and\ \citenamefont
  {Seeger}(2001)}]{Williams01usingthe}%
  \BibitemOpen
  \bibfield  {author} {\bibinfo {author} {\bibfnamefont {Christopher}\
  \bibnamefont {Williams}}\ and\ \bibinfo {author} {\bibfnamefont {Matthias}\
  \bibnamefont {Seeger}},\ }\bibfield  {title} {\enquote {\bibinfo {title}
  {Using the nyström method to speed up kernel machines},}\ }in\ \href@noop {}
  {\emph {\bibinfo {booktitle} {Advances in Neural Information Processing
  Systems 13}}}\ (\bibinfo  {publisher} {MIT Press},\ \bibinfo {year} {2001})\
  pp.\ \bibinfo {pages} {682--688}\BibitemShut {NoStop}%
\bibitem [{Note7()}]{Note7}%
  \BibitemOpen
  \bibinfo {note} {The hat symbol is used for estimators such as the empirical
  eigenvalues, as customary in statistical theory. The hat must not confused
  with the one used for the quantum operators}\BibitemShut {NoStop}%
\end{thebibliography}%

\end{document}